\documentclass{elsart}

\usepackage{amssymb,graphicx}
\newcommand{\beq}{\begin{equation}}
\newcommand{\eneq}{\end{equation}}

\begin{document}

\begin{frontmatter}



\title{Y-junction of superconducting Josephson chains}


\author{Domenico Giuliano$^*$  and Pasquale Sodano$^\dagger$}

\address{$^*$ Dipartimento di Fisica, Universit\`{a} della Calabria and
I.N.F.N., Gruppo collegato di Cosenza, Arcavacata di Rende
I-87036, Cosenza, Italy\\
 $^\dagger$ Perimeter Institute for Theoretical Physics, 31 Caroline
Street, Waterloo, Ontario N2L 2Y5 \thanksref{Perugia}}

\thanks[Perugia]{Permanent address: Dipartimento di Fisica, Universit\`{a} di
Perugia, and I.N.F.N., Sezione di Perugia, Via A. Pascoli, 06123,
Perugia, Italy}

\begin{abstract}

We show that, for pertinent values of the fabrication and control 
parameters,  an attractive finite coupling fixed point emerges 
in the phase diagram of a Y-junction of superconducting Josephson chains.  
The new fixed point arises only when the dimensionless flux $f$ piercing 
the central loop of the network equals $\pi$ and, thus, does not break
time-reversal invariance; for $f \neq \pi$, only the strongly 
coupled fixed point survives as a stable  attractive  fixed point. 
Phase slips (instantons)  have a crucial role in establishing this 
transition: we show indeed that, at
 $f = \pi$, a new set of instantons -the W-instantons- comes 
into play to destabilize the strongly coupled fixed point.
 Finally, we provide a detailed account of the Josephson current-phase 
relationship along the arms of the network,
near each one of the allowed fixed points. Our results evidence 
remarkable similarities between the phase diagram
accessible to a Y-junction of superconducting Josephson chains 
and the one found in the analysis of quantum Brownian
motion on frustrated planar lattices.

\end{abstract}

\begin{keyword}
Wire networks \sep
Phase transitions in model systems \sep Josephson junction arrays
\PACS 71.10.Hf \sep 74.81.Fa \sep   11.25.hf \sep  85.25.Cp
\end{keyword}
\end{frontmatter}


\section{Introduction}

Networks of fermionic and bosonic quantum systems are now attracting
increased attention, due to their relevance to the
engineering of electronic and spintronic nanodevices.  Recently,
in Ref.\cite{aoc}, the transport properties of a Y-junction composed
of three quantum wires enclosing a magnetic flux were studied: modeling the
wires as Tomonaga-Luttinger liquids (TLL), the authors
of Ref.\cite{aoc} were able to show the existence of an attractive
fixed point,  characteristic of the network geometry of the circuit.
A repulsive finite coupling fixed point
has been found in Ref.\cite{Demler}, in the
analysis of Y-junctions of one-dimensional Bose liquids.

Crossed TLL's are the subject of several recent analytical 
\cite{doucosen}, as well as numerical \cite{guowhite} papers: 
these analyses show that, in crossed TLL's, a junction
induces behaviors similar to those arising from impurities 
in condensed matter systems. In Ref.\cite{reyes} it has
been pointed out that, in crossed spin-1/2 Heisenberg chains, 
novel critical behaviors emerge since, as a result
of the crossing, some operators turn from irrelevant to marginal, 
leading to correlation functions exhibiting
power-law decays with nonuniversal exponents.

Impurity models have been largely studied, in connection with
the Kondo models \cite{tscho}, with magnetic chains
\cite{ager}, and for describing static impurities in TLL's
\cite{fencha}.  A renormalization group approach to those systems
leads, after bosonization \cite{gogolin}, to the investigation
of the phases accessible to pertinent boundary sine-Gordon models
\cite{fencha}. Scattering from an impurity often leads the boundary coupling
strength to scale to the strongly coupled fixed
point (SFP),which is rather simple since it describes a fully
screened spin in the Kondo problem or a severed chain in the Kane-Fisher
model \cite{kanefish}. A remarkable exception is provided by
the fixed point attained in the  overscreened Kondo
problem, where an attractive finite coupling fixed point (FFP)
emerges in the phase diagram \cite{tscho}. The FFP is usually
characterized  by novel nontrivial universal indices and by
specific symmetries.

Superconducting Josephson devices  provide remarkable
realizations of quantum systems with  impurities
\cite{giuso1,giuso2}. For superconducting
Josephson chains with an impurity in the middle  \cite{glark,giuso1}
or for SQUID devices \cite{glhek,giuso2} the phase diagram
admits only two fixed points: an unstable weakly
coupled fixed point (WFP), and a stable one at strong coupling.
The boundary field theory approach developed in
Ref.\cite{giuso1,giuso2} not only allows for an accurate
determination of the phases accessible to a superconducting
device, but also  for a field-theoretical treatment of the phase slips
(instantons),  describing quantum tunneling between degenerate
ground-states; furthermore,  it helps to evidence
remarkable analogies with models of quantum Brownian motion on
frustrated planar lattices \cite{kaneyi,saleurb}.

In this paper,  we show that, for pertinent values of the fabrication 
and control parameters, a FFP emerges
in a Y-shaped  Josephson junction  network (YJJN); then, 
we probe the behavior of the YJJN near this fixed point
by computing the Josephson current along the arms of the network. 
The paper is organized as follows:

In section 2 we provide a Luttinger liquid description of the YJJN
and derive the boundary effective Hamiltonian describing the
network;

In section 3 we investigate the fixed points accessible to a YJJN
for different values of the Luttinger parameter $g$ and of the
magnetic field threading the central loop of the YJJN;

Section 4 is devoted to the computation of the current-phase
relation of the Josephson currents along the three arms of the
YJJN, with the purpose of determining the current's pattern near
the fixed points found in section 3. There we evidence the
remarkably different effects of phase slips near the SFP and
the FFP;

In section 5 we argue that -as it happens with other superconducting
devices \cite{shon0} - a YJJN allows to engineer an effective
coherent two-level quantum system, whose states are characterized
by two different macroscopic current's patterns along its arms;

Section 6 is devoted to our concluding remarks, while the appendices
provide the necessary background for the derivation presented  in
the paper.

\section{Effective Hamiltonian of a YJJN}

\begin{figure}
\includegraphics*[width=1.0\linewidth]{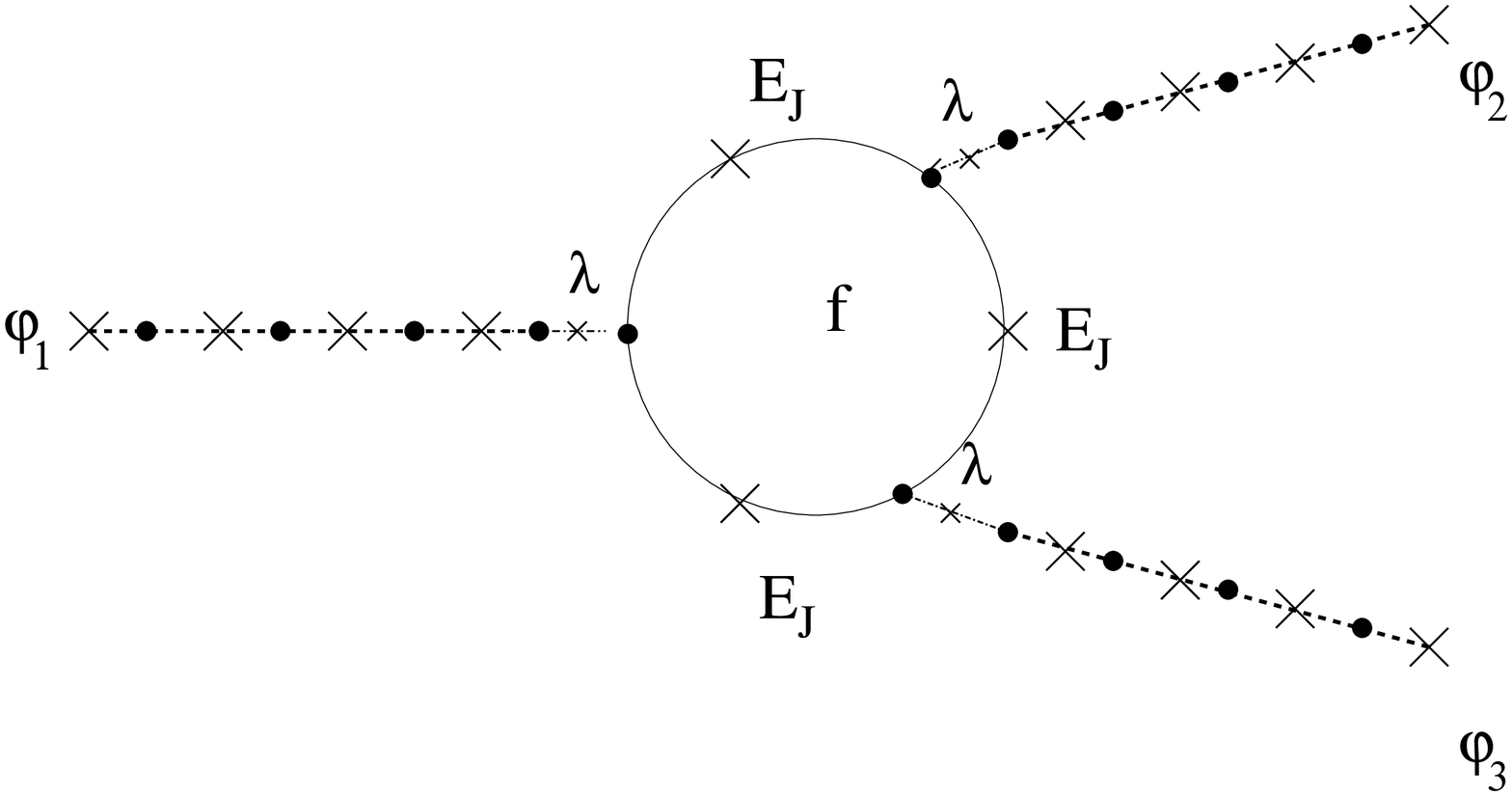}
\caption{$Y$-shaped Josephson junction network: all
the junctions are equal to each other and have nominal Josephson
energy $E_J$, except for the three ones connecting the central region to
the endpoints of the chain, that have nominal energy $\lambda$.}
\label{netw}
\end{figure}

The $Y$-shaped Josephson junction network  we consider is shown in
Fig.\ref{netw}. It is made with three finite Josephson junction (JJ) chains
ending on one side (inner boundary)
with a  weak link  of nominal strength $\lambda$ and on
the other side (outer boundary)
by three bulk  superconductors held at phases $\varphi_j$ ($j=
1,2,3$).  The three chains are connected by the weak links to a circular
JJ chain  {\bf C}, pierced by a dimensionless magnetic  flux $f$.
For simplicity, we assume that all the junctions have Josephson energies $E_J$
and $\lambda \ll E_J$. The Hamiltonian describing the central region,
$H_{\bf C}$, is given by
\beq
H_{\bf C} = \frac{E_c}{2} \sum_{ i = 1}^3 \left[ -i \frac{ \partial }{
\partial \phi_0^{(i)} } - {\bf N'}  \right]^2
- \frac{E_J}{2} \sum_{ i = 1}^3 [ e^{ i [ \phi_0^{(i)} - \phi_0^{(i+1)} +
\frac{f}{3} ]} +  e^{ - i [ \phi_0^{(i)} - \phi_0^{(i+1)} +
\frac{f}{3} ]} ]
\:\:\:\: ,
\label{adi40}
\eneq
\noindent
where $E_c$ is the charging energy of each grain, ${\bf N'}$ is the gate
voltage applied to the i$^{th}$ junction, and $\phi_0^{(i)}$
($i=1,2,3; i+3 \equiv i$) is the phase of the superconducting order parameter
at the $i$-th  grain in {\bf C}.

Following a standard procedure \cite{glark,giuso1,giuso2},
the Hamiltonian in Eq.(\ref{adi40}) can be presented as
\beq
H_{\bf C}  = - \tilde{H} \sum_{ i = 1}^3 [ S_0^{(i)} ]^z
- \frac{E_J}{2} \sum_{ i = 1}^3 \{ e^{ i \frac{f}{3}} [ S_0^{(i)} ]^+
[ S_0^{(i+1)} ]^- +  e^{ - i \frac{f}{3}} [ S_0^{(i+ 1 )} ]^+
[ S_0^{(i)} ]^- \}
\:\:\:\: ,
\label{adi44}
\eneq
\noindent
with  $ \tilde{H} \propto E_c$,
$[ S_0^{(i)} ]^z =  n^{(i)}_0 - {\bf N'} - \frac{1}{2}$ and
$[ S_0^{(i)} ]^\pm = e^{ \pm i \phi_0^{(i)} }$,
where $n^{(i)}_0$ is the total charge at grain $i$ (measured in units of
$e^*$).

For $\tilde{H} > E_J >0$, the eigenstates of Eq.(\ref{adi44}) are
given by

\begin{itemize}

\item A ``fully polarized'' ground state:

 $ | 0 \rangle =
| \uparrow \uparrow \uparrow \rangle$, with energy $\epsilon_0 =
-\frac{3}{2} \tilde{H}$;
\item A low-energy triplet of states:

$| 1 , 1 \rangle = \frac{1}{\sqrt{3}} [ | \downarrow \uparrow \uparrow
\rangle + | \uparrow \downarrow \uparrow  \rangle + |  \uparrow \uparrow
\downarrow \rangle  ] $,
with energy $\epsilon_{1,1} ( f ) =
- \frac{ \tilde{H} }{2} - E_J \cos ( \frac{ f}{3} )$;

$ | 1 , 2 \rangle = \frac{1}{\sqrt{3}} [ | \downarrow \uparrow \uparrow
\rangle - e^{ - i \frac{ \pi}{3}}
| \uparrow \downarrow \uparrow  \rangle  - e^{  i \frac{ \pi}{3}}
|  \uparrow \uparrow \downarrow \rangle  ]$, with energy  $\epsilon_{1,2}
( f ) = - \frac{ \tilde{H} }{2} - E_J \cos ( \frac{ f - \pi }{3} )$;

$| 1 , 3 \rangle = \frac{1}{\sqrt{3}} [ | \downarrow \uparrow \uparrow
\rangle - e^{  i \frac{ \pi}{3}}
| \uparrow \downarrow \uparrow  \rangle  - e^{ - i \frac{ \pi}{3}}
|  \uparrow \uparrow \downarrow \rangle  ]$,
with energy $ \epsilon_{1,3} ( f ) =
- \frac{ \tilde{H} }{2} - E_J \cos ( \frac{ f + \pi }{3} ) $;
\item A high-energy triplet of states:

$| 2 , 1 \rangle = \frac{1}{\sqrt{3}} [ | \downarrow \downarrow \uparrow
\rangle + | \downarrow \uparrow \downarrow  \rangle + |  \uparrow \downarrow
\downarrow \rangle  ]$,
with energy $ \epsilon_{2,1} ( f ) =
 \frac{ \tilde{H} }{2} - E_J \cos ( \frac{ f }{3} )$;

$| 2 , 2 \rangle = \frac{1}{\sqrt{3}} [ | \uparrow \downarrow \downarrow
\rangle - e^{  i \frac{ \pi}{3}}
| \downarrow \uparrow \downarrow  \rangle  - e^{  - i \frac{ \pi}{3}}
|  \downarrow \downarrow  \uparrow \rangle  ]$,
with energy $ \epsilon_{2,2} ( f ) = \frac{ \tilde{H} }{2}
 - E_J \cos ( \frac{ f - \pi }{3} )$;

$| 2 , 3 \rangle = \frac{1}{\sqrt{3}} [ | \uparrow \downarrow \downarrow
\rangle - e^{  - i \frac{ \pi}{3}}
| \downarrow \uparrow \downarrow  \rangle  - e^{  i \frac{ \pi}{3}}
|  \downarrow \downarrow
\uparrow \rangle  ]$, with energy $\epsilon_{3,2} ( f ) =
\frac{ \tilde{H} }{2} - E_J \cos ( \frac{ f+ \pi }{3} )$;
\item A high-energy fully-polarized state $| 3 \rangle =
| \downarrow \downarrow \downarrow \rangle$, with energy
$\epsilon_3 = 3 \tilde{H}$.
\end{itemize}
We require that {\bf C} is connected to the three finite chains via a
charge tunneling Hamiltonian $H_T$, given by

\beq
H_T = -  \lambda    \sum_{ i = 1}^3 \cos [ \phi_1^{(i)} - \phi_0^{(i)}  ]
\;\;\;\; .
\label{adi53}
\eneq
\noindent
Since $\lambda / E_J \ll 1$, one may resort
to a  Schrieffer-Wolff transformation \cite{swolff}, to derive an
Hamiltonian $H_B$ describing the effective boundary interaction at the
inner boundaries of the three chains. To the second order in $\lambda$,
$H_B$ is given by
\beq
H_B \approx B ( f ) \sum_{ i = 1}^3
e^{ - i \phi_1^{(i)}}  e^{  i \phi_1^{(i)} } + A ( f )  \sum_{ i = 1}^3
 e^{  - i \phi_1^{(i)}}   e^{   i \phi_1^{(i+1)} } + {\rm h. c.}
\;\;\;\; ,
\label{adi54}
\eneq
\noindent
where $ B ( f ) =  \frac{\lambda^2}{ 12} \sum_{k=1}^3
\left(  \frac{ 1 }{ \epsilon_0 - \epsilon_{ 1 , k } ( f ) } \right)$, and
$ A ( f ) =   \frac{\lambda^2}{ 12} \sum_{k=1}^3
\left(  \frac{  e^{- \frac{2}{3} \pi i
(k-1)}  }{ \epsilon_0 - \epsilon_{ 1 , k } ( f ) } \right)$.

$ A ( f )$ is, in general, a complex number, equal to $-E_W e^{ i \gamma}$
($E_W > 0 $). Its phase $\gamma$ is related to the magnetic flux by

\beq
\tan \gamma = \frac{ \sqrt{3}}{2} \left[
\frac{ \sum_{ k = 2, 3} \frac{ (-1)^k}{ \epsilon_0 - \epsilon_{1,k} ( f )}
}{ \frac{1}{\epsilon_0 - \epsilon_{1,1}
( f )} - \frac{1}{2}  \sum_{ k = 2, 3} \frac{ 1}{
\epsilon_0 - \epsilon_{1,k} ( f )} } \right]
\:\:\:\: ;
\label{psc1}
\eneq
\noindent
for $f = 2 k \pi $ and $( 2 k +1 ) \pi $ ($k= 0 , \pm 1 , \pm 2 , \ldots$),
$\gamma = 2 k \pi / 3 $ and $ ( 2 k +1 ) \pi / 3$, respectively.

The Hamiltonian describing the three finite chains may be written as
\cite{glark}

\[
H_{0} = \frac{E_c}{2} \sum_{ i = 1,2,3}
\sum_{ j = 1 }^{ L / a } \left[ - i \frac{
\partial }{ \partial \phi_j^{(i)}  }-  {\bf N} \right]^2
+
\]
\beq
 \sum_{ i = 1,2,3} \sum_{ j = 1}^{ L/a - 1 } \left[  - E_J
\cos ( \phi_j^{(i)}  - \phi_{ j + 1 }^{(i)}  ) +  E^z \left(  - i \frac{
\partial }{ \partial \phi_j^{(i)}  }-  {\bf N} \right)\left(
- i \frac{ \partial }{ \partial \phi_{j+1}^{(i)} }- {\bf N} \right) \right]
\:\:\:\: .
\label{eql.1}
\eneq
\noindent
In Eq.(\ref{eql.1}) $\phi_j^{(i)}$ is the phase of the superconducting
order parameter at grain $j$ of the $i$th chain,
$- i \frac{ \partial }{ \partial \phi_j^{(i)} }$ 
is the corresponding charge operator;   ${\bf N}$ is proportional to the gate
voltage  $V_g$ applied to each grain, while $L$ and $a$ are the length
of each  chain and the lattice spacing, respectively. $E^z$
accounts for  the Coulomb repulsion between charges on nearest
neighboring junctions.  Following the procedure detailed in appendix
A, Eq.(\ref{eql.1})  may be written in Tomonaga-Luttinger (TL)-form
\cite{luttinger} as

\beq
H_0 = \sum_{ j =1, 2, 3} \frac{g}{ 4 \pi} \: \int_0^L \: d x \: \left[
\frac{1}{v} \left( \frac{ \partial \Phi_j }{ \partial t} \right)^2
+ v  \left( \frac{ \partial \Phi_j }{ \partial x} \right)^2 \right]
\;\;\;\; ,
\label{adi54.b}
\eneq
\noindent
where the fields
$\Phi_j ( x )$ ($j=1,2,3$) describe the collective plasmon modes
of the chains,  $\Delta = E^z- \frac{3}{16} \frac{(E_J)^2}{E_c}$,
 $v = v_f \sqrt{1 +  \frac{ 4 \pi a \Delta
[ 1 - \cos ( 2 a k_f )]}{v_f} }$, and $g = \sqrt{\frac{v_f}{v_f +  4 \pi a
\Delta [ 1 - \cos ( 2 a k_f )]} }$, with $v_f = 2 \pi E_J \sin ( a k_f )$
and $k_f = {\rm arccos} ( h E_c / E_J )$.

Since at the outer boundary the three chains are connected to
three bulk superconductors at fixed phases $\varphi_j$, the fields
$\Phi_j$ must satisfy  the Dirichlet boundary conditions
\beq
\Phi_j ( L ) = \sqrt{2}  [ 2 \pi n_j + \varphi_j ],
\label{bea}
\eneq
\noindent
 where $j=1,2,3$ and $n_j$ are integers. On the
inner boundary, the three chains  are connected to  {\bf C} via $H_T$: as
a result, one should impose here Neumann boundary conditions (i.e.,
$ \frac{ \partial \Phi_j ( 0 )}{ \partial x} = 0 $). For our following
analysis, it is most convenient to introduce
linear combinations of the plasmon fields, such as
$X ( x ) = \frac{1}{ \sqrt{3}} \sum_{ j = 1}^3 \Phi_j ( x )$,
$\chi_1 ( x ) = \frac{1}{ \sqrt{2}} [ \Phi_1 ( x ) - \Phi_2 ( x ) ] $, and
$\chi_2 ( x ) = \frac{1}{ \sqrt{6}} [ \Phi_1 ( x ) + \Phi_2 ( x ) - 2 \Phi_3
(x ) ]$. Since $E_W$ is of order $\lambda^2 / E_J$, one has that
$E_W / E_J \ll 1$ and, thus, at $x=0$, the fields $\chi_1 , \chi_2$
also satisfy Neumann boundary conditions. Of course, at the outer boundary,
$\chi_1 , \chi_2$ satisfy Dirichlet boundary conditions.

In the long wavelength limit, the first term on the r.h.s. of
Eq.(\ref{adi54}) may be well approximated  as  $3 B ( f ) +
 { \rm const}  \frac{ \partial X ( 0 )}{ \partial x}$. Due to
Neumann boundary conditions, this term only contributes by an
irrelevant constant to Eq.(\ref{adi54}). Eq.(\ref{adi54}) may be,
then, usefully presented in the form

\beq
H_B = - 2 \bar{E}_W \sum_{ i = 1}^3 : \cos [ \vec{\alpha}_i
\cdot \vec{\chi} ( 0 )  + \gamma ]: \;\;\;\; ,
\label{let3}
\eneq
\noindent
with $\vec{\alpha}_1 = ( 1 , 0 )$,
$\vec{\alpha}_2 = ( - \frac{1}{2} , \frac{ \sqrt{3}}{2} )$,
$\vec{\alpha}_3 = ( - \frac{1}{2} , - \frac{ \sqrt{3}}{2} )$.
The  colons ($: \: :$) in  Eq.(\ref{let3})  denote normal
ordering with respect to the vacuum of the bosonic fields
$\chi_1 , \chi_2$. The effective coupling $\bar{E}_W$ is given by $\bar{E}_W =
\left( \frac{a}{L} \right)^\frac{1}{g} E_W$.
Eq.(\ref{let3}) may be regarded as  the bosonic version of the
boundary Hamiltonian describing the central region of a $Y$-junction
of three quantum wires, introduced in Ref.\cite{aoc}.  As we shall
see, setting $\gamma = ( 2 k + 1 ) \pi / 3 $, allows for the emergence of a new
attractive fixed point also in the phase diagram of the $Y$-junction of superconducting Josephson chains.
It should be noticed that
this fixed point is attractive, since
in a superconducting network, bosons are charged; this should be
contrasted  with the situation arising in  $Y$-shaped
networks of neutral atomic condensates \cite{Demler}, where the
FFP is repulsive.

\section{Phase diagram  of a YJJN}

In this section, we use the renormalization group approach to
investigate the phases accessible to a superconducting
YJJN. As evidenced in the analysis of
other superconducting devices \cite{giuso1,giuso2}, there is usually
a range of values of the Luttinger parameter $g$ for which the
phase diagram allows for a crossover from an unstable WFP to a stable
SFP.  For a Josephson chain with an impurity \cite{glark,giuso1} and
for SQUID devices \cite{glhek,giuso2}, the crossover is driven by
the ratio $L / L_*$, where $L$ is the length of
the chain (or the diameter of the superconducting loop in a SQUID) and $L_*$ is a pertinently defined healing length \cite{giuso1}.
Here, we shall show that,  when $\gamma = (2 k + 1 ) \pi / 3$, a new
relevant boundary interaction, emerging in a YJJN  at strong coupling,
destabilizes the SFP: as a result, since the
WFP is IR unstable, an IR stable attractive FFP emerges in the phase diagram.
Remarkably, for these values of $\gamma$,
the phase diagram of a YJJN is similar to the one accessible to
a bosonic quantum Brownian particle on planar frustrated lattices
\cite{kaneyi}, and to spin-1/2 fermions hopping on $Y$-junctions
of quantum wires  \cite{aoc}.

\subsection{The weakly coupled fixed point}

Setting $\bar{E}_W =0$ defines the WFP, where
the fields $\chi_1 ( x ) , \chi_2 ( x )$  obey Dirichlet boundary conditions
at the outer boundary and Neumann boundary conditions at the
inner boundary. As a result, the  mode expansion of $\chi_i$ is given by

\beq
\chi_i ( x , t ) = \xi_i + \sqrt{\frac{2}{g}} \: \sum_n \cos \left[
\frac{ \pi}{L} \left( n + \frac{1}{2} \right) x \right]
\frac{ \alpha_i ( n ) }{ n + \frac{1}{2}} \: e^{ - i \frac{\pi}{L} \left( n +
\frac{1}{2} \right) v t }
\:\:\:\: ,
\label{cou2}
\eneq
\noindent
with $  [ \alpha_i ( n ) , \alpha_j ( n' ) ] = \delta_{ij}
\left( n + \frac{1}{2} \right) \delta_{ n +n' - 1, 0}$,
$\xi_1 =   \mu_1   + 2 \pi n_{12}$,
$\xi_2 = \mu_2 +  \frac{2}{ \sqrt{3}} \left[ 2 \pi n_{13}
- \pi n_{12} \right]$, $ ( \mu_1 , \mu_2 ) =
( [ \varphi_1 - \varphi_2 ] ,   \frac{2}{ \sqrt{3}} [ (
\varphi_1 - \varphi_3 ) - (\varphi_1 - \varphi_2 ) / 2 ] )$,
 with $n_{ij}=n_i-n_j$.

The perturbative renormalization
group equations may be derived from the partition function, written as
a power series in the boundary interaction strength. From Eq.(\ref{let3}),
one gets

\[
\frac{ {\bf Z} }{ {\bf Z}_0 }
= \sum_{ N = 0}^\infty \bar{E}_W^N \sum_{\epsilon_1 , \ldots ,
\epsilon_N = \pm 1} \: \exp \left[ \sum_{j=1}^N \epsilon_{j} \gamma \right]
\times
\]
\beq
\int_0^\beta \: d \tau_1
\: \int_0^{\tau_1 - \frac{a}{v}} \: d \tau_2 \: \ldots \:
 \int_0^{\tau_{N-1} - \frac{a}{v}} \: d \tau_N \:
\langle {\bf T}_\tau \prod_{j =1}^N  : \exp [ i \epsilon_j
\vec{\alpha}_{k_j} \cdot \vec{\chi} ( \tau_j ) ] : \rangle_0
\:\:\:\: ,
\label{cough1}
\eneq
\noindent
with ${\bf Z}_0 = \prod_{ n = 0}^\infty [ 1 -
\bar{q}^{ n + \frac{1}{2} } ]^2 $,
$\bar{q} = \exp \left[ - \beta \frac{ \pi v  }{L} \right]$), and
$\beta = ( k_B T )^{-1}$. In Eq.(\ref{cough1}), the  lattice step $a$ has to
be regarded  as the short-distance cutoff, $\langle \ldots \rangle_0$
denotes thermal averages with respect to ${\bf Z}_0$, and
${\bf T}_\tau$ denotes imaginary time ordered products.

The $N$-point functions of the vertex operators  $ : \exp [ i \epsilon_j
\vec{\alpha}_{k_j} \cdot \vec{\chi} ( \tau_j ) ] :$ are readily computed
using Wick's theorem for vertex operators  \cite{pginsp}. As $\beta v 
/ L \gg 1$, they are given by

\beq
\langle {\bf T}_\tau \prod_{j =1}^N  : \exp [ i \epsilon_j
\vec{\alpha}_{k_j} \cdot \vec{\chi} ( \tau_j ) ] : \rangle_0 =
 \exp \left[ \frac{2}{g} \sum_{ i < j = 1}^N \epsilon_i \epsilon_{j}
 \vec{\alpha}_{k_i} \cdot  \vec{\alpha}_{k_{j}} \:
\gamma_\tau ( \tau_i , \tau_j ) \right]
\:\:\:\: ,
\label{cou7}
\eneq
\noindent
with
\beq
\gamma_\tau ( \tau , \tau' ) = \ln \left| \frac{ e^{ \frac{ \pi}{2 L }v \tau}
-  e^{ \frac{ \pi}{2 L } v \tau'} }{ e^{ \frac{ \pi}{2 L } v \tau}
+  e^{ \frac{ \pi}{2 L } v \tau'} } \right|
\:\:\:\: .
\label{cou8}
\eneq
\noindent
As a result, at the WFP, one sees that the  scaling dimension
of the boundary interaction in Eq.(\ref{let3}) is given by $h_W ( g ) =
\frac{1}{g}$, and  that the dimensionless  coupling strength $G ( L ) =
L \bar{E}_W$ scales as  $G ( L ) \sim L^{1-\frac{1}{g}}$.

From the operator product expansion (O.P.E.)  between  vertex operators
\beq
\left\{ : \exp \left[ i \vec{\alpha}_i \cdot \chi ( \tau ) \right] :
: \exp \left[ i \vec{\alpha}_j \cdot \chi ( \tau' ) \right] : \right\}_{ \tau'
\to \tau^- } \approx \left[ \frac{v ( \tau - \tau' )}{ L} \right]^{
- \frac{2}{g}} :  \exp \left[ - i \vec{\alpha}_k \cdot \vec{\chi} ( \tau )
\right]:
\;\;\;\; ,
\label{jlu1}
\eneq
\noindent
with $i \neq j \neq k$, one gets the second-order
renormalization group  equations for the complex coupling $G ( L ) e^{
i \gamma} $ as
\beq
\frac{ d [ G ( L) e^{ i \gamma} ] }{ d \ln ( \frac{L}{L_0} ) } =
[ 1 - \frac{1}{g}  ] [ G ( L ) e^{ i \gamma}  - 2 G^2 ( L ) e^{ - 2 i \gamma}
\;\;\;\; ,
\label{ju2}
\eneq
\noindent
which may be usefully presented  as
\begin{eqnarray}
\frac{ d G ( \ell )}{ d \ell} &=& [ 1 - \frac{1}{g}  ] G ( \ell ) +
2 \cos ( 3 \gamma )  G^2 ( \ell ) \label{cou19.a} \\
\frac{ d \gamma}{ d \ell} &=& - 2 \sin ( 3 \gamma) G^2 ( \ell )
\:\:\:\:
\label{cou19}
\end{eqnarray}
\noindent
( $\ell = \ln \left( \frac{L}{L_0} \right)$).
Since Eqs.(\ref{cou19.a},\ref{cou19})  are periodic under
$\gamma \longrightarrow \gamma + \frac{2 \pi}{3}$, the
resulting phase diagram of the YJJN will present the same
periodicity. Also, the phase  diagram strongly depends on
whether $g<1$, or $g>1$. Indeed:

\begin{enumerate}

\item For  {\bf $g< 1$},  the linear term in  Eq.(\ref{cou19.a}) has
a negative coefficient and, thus,  $\forall \gamma $,  the system is
attracted by a fixed point with $G_* = 0$.
Furthermore, Eq.(\ref{cou19}) shows that the value of $\gamma$ at the
attractive fixed point is $\gamma_* = 2 k \pi / 3$,
if $    ( 2 k - 1 ) \pi / 3  < \gamma (L_0 ) <  ( 2 k + 1 ) \pi $,
while it is $\gamma^* =  ( 2 k + 1 ) \pi / 3$ if $\gamma ( L_0 )
= ( 2 k + 1 ) \pi / 3$ \footnote{$\gamma ( L_0 )$ is the value of the phase
$\gamma$ at the reference length $L_0$. It should be noticed that, if
 $\gamma ( L_0 ) =  ( 2 k + 1 ) \pi / 3$, $\gamma$ does not scale with $L$.}.

\item For {\bf $g> 1$}, Eq.(\ref{cou19.a}) has a positive coefficient;
as a result, $G ( \ell )$ grows as $\ell$ increases.  Whether $G_*$ is now
finite, or $\infty$, depends on the values of $g$ and $\gamma (L_0 )$.

\end{enumerate}

In the following subsection, we will derive the perturbative
RG equations near the SFP. We  shall
see that, for $g > \frac{9}{4}$ (and for any value of $\gamma ( L_0 )$) ,
the system is attracted by a fixed point with $G_* = \infty$.
For $1 < g < \frac{9}{4}$ and for $\gamma ( L_0 ) =
( 2 k + 1 ) \pi / 3$, the SFP becomes unstable since, for $\gamma ( L_0 ) =
( 2 k + 1 ) \pi / 3$, a new leading boundary perturbation
arises at the SFP. As a consequence
a stable attractive fixed point emerges in the phase diagram at a finite
value of $G_*$. It is easy to convince oneself
that, for $1 < g < \frac{9}{4}$ and for
$\gamma ( L_0 ) \neq   ( 2 k + 1 ) \pi / 3$, the stable fixed point is still
at  $G_* = \infty$.

\subsection{The strongly coupled fixed point}

The SFP is reached when the running
coupling constant  $G$ goes to $\infty$. The fields $\chi_j ( x )$, $j=1,2$,
now obey Dirichlet boundary conditions at $x=0$. The allowed values of
$\chi_1 ( 0 ) , \chi_2 ( 0 )$ are determined by  the manifold of
the minima of the effective boundary potential  (Eq.(\ref{let3})).
It is easy to see that:
\begin{enumerate}

 \item for  $ ( 6 k  - 1 ) \pi / 3 <  \gamma <
 ( 6 k  + 1 ) \pi / 3 $, the minima lie on
the triangular sublattice A, defined by $ (\chi_1 (0)  , \chi_2  ( 0 ) ) =
( 2 \pi n_{12}  , \frac{2}{ \sqrt{3}} [ 2 \pi n_{13}  + \pi n_{12} ] ) $.

\item for $ ( 6 k  + 1 ) \pi / 3 <  \gamma <
 ( 6 k  + 3 ) \pi / 3 $, the minima lie on the triangular sublattice B, given by
$ (\chi_1 (0), \chi_2 (0) ) = ( 2 \pi n_{12} + \frac{4 \pi}{3} ,
\frac{2}{ \sqrt{3}} [ 2 \pi n_{13}  + \pi n_{12}  ] )$.

\item for $ ( 6 k  + 3 ) \pi / 3 <  \gamma <
 ( 6 k  + 5 ) \pi / 3 $, the minima lie on the triangular sublattice C, given by
$ (\chi_1 (0), \chi_2 (0) ) = ( 2 \pi n_{12} - \frac{4 \pi}{3} ,
\frac{2}{ \sqrt{3}} [ 2 \pi n_{13}  + \pi n_{12}  ] )$.

\end{enumerate}
At $\gamma =  ( 6 k  + 1 ) \pi / 3 $,  $\gamma =  ( 6 k  + 3 ) \pi / 3 $,
 $\gamma =  ( 6 k  + 5 ) \pi / 3 $, the two sublattices A and B, B and C, and
C and A become degenerate in energy, respectively.  From Eq.(\ref{let3}),
one sees that, for $\gamma \sim ( 6 k  + 1 ) \pi / 3 $, the  difference
in energy between the sets of the minima forming the A and B
sublattices is given by $\sim \bar{E}_W \sin \left[ \gamma -
\frac{(2k+1) \pi}{3} \right]$.
Similar expression hold for the difference
in energy between the sets of the minima forming the B and C sublattices
for  $\gamma =  ( 6 k  + 3 ) \pi / 3 $, and for the difference
in energy between the sets of the minima forming the C and A sublattices
for  $\gamma =  ( 6 k  + 5 ) \pi / 3 $.

The Dirichlet boundary conditions at both boundaries are consistent with
the mode expansions

\beq
\chi_j ( x , t ) = \xi_j +
\sqrt{ \frac{2}{g}} \left\{ - \frac{  \pi x}{L} P_j - \sum_{ n \neq 0}
\sin \left[ \frac{ \pi n x}{L} \right] \frac{ \alpha_n^j}{n}
e^{ - i \frac{\pi}{ L} n v t } \right\}
\;\;\;\; ,
\label{scou1}
\eneq
\noindent
with $ [ \alpha_n^i , \alpha_m^j ] = \delta^{i,j} n \delta_{m+n , 0 }$.

For  $ ( 6 k  - 1 ) \pi / 3 <  \gamma < ( 6 k  + 1 ) \pi / 3 $,
the eigenvalues of the zero-mode operators $P_j$ are proportional to the
coordinates of the sites of the sublattice A,
and are given by

\beq
(p_1 , p_2 )_A =  \sqrt{ 2 g  }
\left( \left[  n_{12} + \frac{\mu_1}{ 2 \pi} \right] , \left[
 \frac{ \mu_2 }{2 \pi }  +  \frac{ 2 }{ \sqrt{3} } \left(
n_{13} +  \frac{n_{12}}{2}  \right) \right] \right)
\:\:\:\: ;
\label{subla}
\eneq
\noindent
for  $ ( 6 k  + 1 ) \pi / 3 <  \gamma < ( 6 k  + 3 ) \pi / 3 $, they
are proportional to the coordinates of the sites of the sublattice B,
and are given by

\beq
(p_1 , p_2 )_B  = \sqrt{  2 g  }  \left( \left[ n_{12} +
\frac{ \mu_1 }{ 2 \pi } + \frac{ 2}{3} \right] , \left[
\frac{ \mu_2 }{ 2 \pi}
+ \frac{ 2}{ \sqrt{ 3} } \left(  n_{13} + \frac{ n_{12}}{2}
\right) \right] \right)
\:\:\:\: ;
\label{sublb}
\eneq
\noindent
finally, for  $ ( 6 k  + 3 ) \pi / 3 <  \gamma < ( 6 k  + 5 ) \pi / 3 $, they
are proportional to the coordinates of the sites of the sublattice C,
and are given by

\beq
(p_1 , p_2 )_C  = \sqrt{  2 g  }  \left( \left[ n_{12} +
\frac{ \mu_1 }{ 2 \pi } - \frac{ 2}{3} \right] ,
\left[ \frac{ \mu_2 }{ 2 \pi}  +
 \frac{ 2}{ \sqrt{ 3} } \left(  n_{13} + \frac{ n_{12} }{2}
 \right) \right] \right)
\:\:\:\: .
\label{sublc}
\eneq
\noindent
The eingenstates associated to the above eigenvalues shall be denoted as $ | n_{12},n_{13} \rangle_ \ell $ where $\ell=A,B,C$.
At the degeneracy points, the merging of two sublattices of minima
implies a merging of the lattices of eigenvalues of the zero-mode
operators: for instance, for $\gamma = \pi / 3$ the set of the
allowed eigenvalues of $ ( P_1 , P_2 )$ contains both the values
$ (p_1 , p_2 )_A$ and $ (p_1 , p_2 )_B$, for $\gamma = \pi$, it
 contains both the values $ (p_1 , p_2 )_B$ and $ (p_1 , p_2 )_C$,
for $\gamma = 5 \pi / 3$, it
contains both the values $ (p_1 , p_2 )_C$ and $ (p_1 , p_2 )_A$.

At the SFP, one may separately  compute the contribution of any one
of the sublattices A, B and C to the total
partition function as

\beq
{\bf Z}_\ell [ \vec{\mu} ] = \frac{1}{\eta^2 ( \bar{q} )} 
\sum_{ n_{12} , n_{13} \in Z} \exp \left\{ - \beta \frac{ \pi v g}{ L} \left[
\left( n_{12} + \frac{\mu_1}{2 \pi} + \frac{ 2 \epsilon_\ell}{3} \right)^2
+ \frac{4}{3} \left( n_{13} + \frac{n_{12}}{2}  +
\frac{\sqrt{3}}{4 \pi} \mu_2 \right)^2 \right] \right\}
\: .
\label{cou4}
\eneq
\noindent
In Eq.(\ref{cou4}),  $\ell = A , B , C$,  $\epsilon_A = 0,
\epsilon_B = 1 , \epsilon_C = -1$, while $\eta ( x ) = \prod_{n=1}^\infty
( 1 - x^n)$, and $\bar{q}$ has been defined after Eq.(\ref{cough1}).

If one  denotes by $\psi_1 , \psi_2$ the  fields dual to $\chi_1$ and $\chi_2$,
 one may easily write their mode expansion as

\beq
\psi_j ( x , t ) = \sqrt{2g} \left\{ \theta_0^j + \frac{  \pi v t }{L} P_j
+ i \sum_{ n \neq 0} \cos \left[ \frac{ \pi n x}{L} \right]
\frac{ \alpha_n^j}{n}  e^{ - i \frac{ \pi }{L} n v t } \right\}
\:\:\:\: ,
\label{scou2}
\eneq
\noindent
with $[ \theta_0^j , P_i ] = i \delta_{i , j } $, and $j=1,2$.

For  $ \gamma \neq ( 2 k + 1 ) \pi /3$, the minima of the boundary
potential span only one of the sublattices A, B and C. In this
case,   the leading boundary perturbation at the inner
boundary is given by a linear combination of the dual vertex operators
$\tilde{V}^\pm_1$,  $\tilde{V}^\pm_2$, and   $\tilde{V}^\pm_3$, defined in
terms of the dual fields as

\beq
\tilde{V}^\pm_j = : \exp \left[ \pm i 2 \sqrt{\frac{2}{3} }
\vec{\rho}_j \cdot \vec{\psi} ( 0 ) \right] :
\:\:\: , \;\;   (j=1,2,3) \:\;\;\; ,
\label{cou5}
\eneq
\noindent
with $\vec{\rho}_1 = ( 0 , 1 )$,  $\vec{\rho}_2 = ( \frac{\sqrt{3}}{2} ,
- \frac{1}{2}  )$,  $\vec{\rho}_3 =  ( - \frac{\sqrt{3}}{2} ,
- \frac{1}{2}  )$:
they describe instanton trajectories connecting two sites in
one  of the triangular sublattices A, B or C (``V-instantons'').
The two-point correlation  function of the dual boundary vertices is given by
\begin{figure}
\includegraphics*[width=1.0\linewidth]{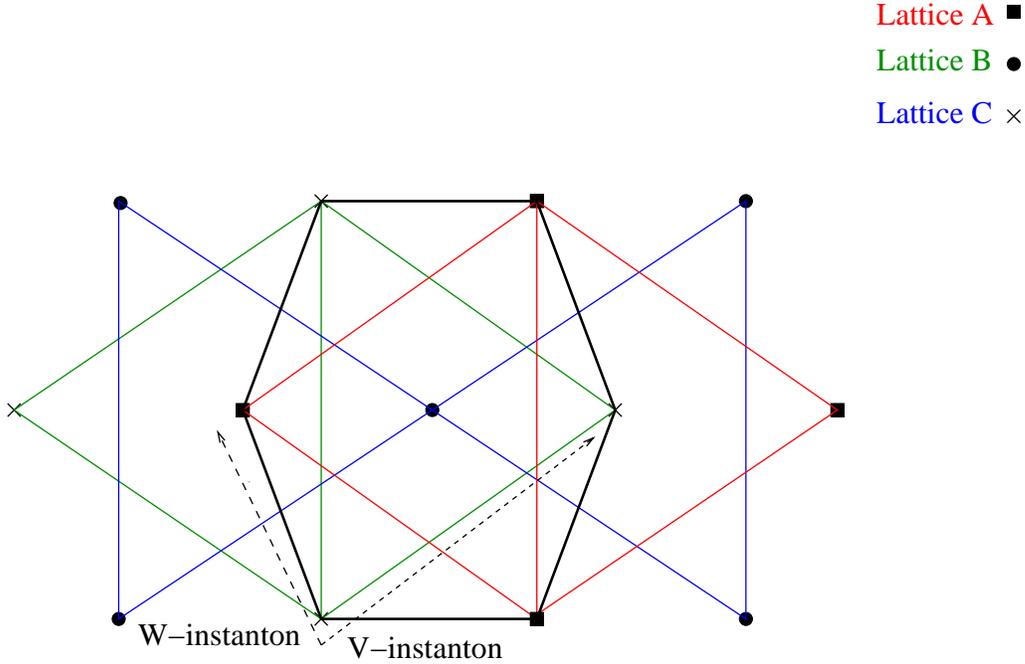}
\caption{Points on the three triangular  sublattices A, B and C: at
$\gamma = ( 2 k + 1 ) \pi / 3$, the energies of two sublattices are degenerate
and the  minima of the boundary potential span a
honeycomb lattice, whose sites are
connected  by W-instanton trajectories, shorter than the
V-instanton trajectories, connecting sites on the same
sublattice. The black honeycomb is an elementary cell of the lattice of the
minima emerging at  $ \gamma = \pi / 3$. }
\label{ablat}
\end{figure}

\beq
\langle  \tilde{V}^\pm_j ( \tau )  \tilde{V}^\mp_i ( \tau' )  \rangle
\propto \delta_{j,i} \left[ e^{ \frac{ \pi v \tau}{L} }  -
e^{ \frac{ \pi v \tau'}{L} } \right]^{ - \frac{8g}{3} }
\:\:\:\: ,
\label{cou7a}
\eneq
\noindent
and, thus, the scaling dimension of  $\tilde{V}^\pm_j ( \tau ) , j=1,2,3$,
is given by  $h_S ( g ) = \frac{4 g}{3}$. As a result,  the SFP
 is stable for $g > 3 / 4$ and for $\gamma \neq ( 2 k + 1 ) \pi / 3$.
Thus, for $3 /4 < g < 1$  and for $\gamma \neq ( 2 k + 1 ) \pi / 3$,
 both the WFP, and the SFP are stable and,
accordingly, the phase diagram allows for a repulsive FFP.
For $g > 1$ and for $\gamma \neq ( 2 k + 1 ) \pi / 3$,
the SFP is the only IR stable fixed point:  the set of the allowed
eigenvalues of $ ( P_1 , P_2 )$ depends upon the value of $\gamma$, as
discussed above. Accordingly, the fixed point partition function is given by
${\bf Z}_\ell [ \vec{\mu} ]$  in Eq.(\ref{cou4}), for a pertinent
choice of $\ell$. Remarkably, this shows that the SFP is time-reversal
invariant, even if the ``bare'' value of $\gamma$ breaks this symmetry.
This is not surprising, though, as the symmetry of the system at the
IR stable fixed-point is usually higher than the symmetry
of the microscopic system.

For $\gamma = ( 2 k + 1 ) \pi/3$,   the  sets of minima belonging to
 two sublattices have the same energy.
As a result, the  eigenvalues of the zero-mode
operators lie all on a honeycomb lattice obtained by  merging  two
triangular sublattices, as sketched in Fig.\ref{ablat} for 
$\gamma = \pi / 3$, at which  point the sublattices A 
and B merge into a honeycomb lattice.
The leading perturbation  near the Dirichlet fixed point contains, now,
operators representing ``shorter'' jumps between neighboring
minima on the honeycomb lattice (``W-instantons'').

\begin{figure}
\includegraphics*[width=1.0\linewidth]{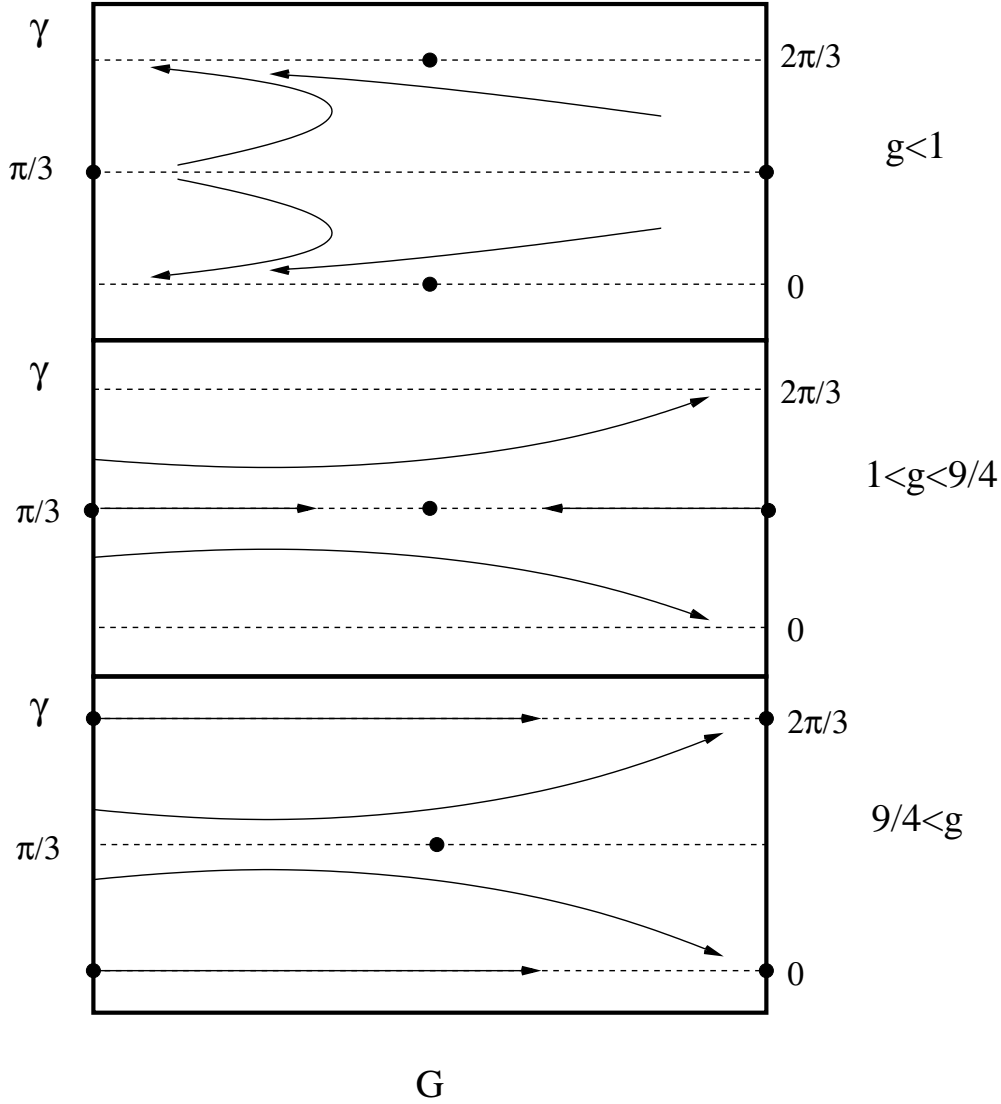}
\caption{Phase diagram of a YJJN  for (from top to bottom):
$g<1$ (the only IR stable fixed point is at
$G_*=0$); $1<g<9/4$ (the  IR stable fixed point is  either at
$G_*=\infty$, or at a finite $G_*$, according to the
initial value of $\gamma$); $9/4<g$  (the only fixed point is at
$G_* = \infty$).}
\label{jla}
\end{figure}

Following   Ref.\cite{kaneyi}, one may describe these instantons by
introducing an  isospin operator
$\vec{\tau}$, acting on a pertinent    two-component spinor \footnote{An
 $\uparrow$-spinor is associated to a minimum
lying on sublattice A and  a $\downarrow$ spinor to a minimum
lying on sublattice B.}.
As a result, the leading boundary perturbation may now be written as

\beq
H_B  = - \xi \sum_{ i = 1}^3 \{ \tau^+ W_i^\dagger ( \tau ) +
\tau^- W_i ( \tau ) \}
\:\:\:\: ,
\label{abe42}
\eneq
\noindent
with $W_j ( \tau ) = : \exp \left[ \frac{2}{3} i \vec{\alpha}_j
\cdot \vec{\psi} ( \tau ) \right] :$ and $\xi \sim E_J - E_W$.
Since the boundary interaction contains the
isospin operators $\vec{\tau}$, the relevant O.P.E.'s are
obtained by combining the multiplication rules for the isospin operators

\beq
\tau^z \tau^\pm = \pm \tau^\pm \;\;\; ; \;\;
\tau^\pm \tau^\mp = 1 \pm \tau^z \;\;\:\: ,
\label{ju3}
\eneq
\noindent
with the O.P.E.'s of the bosonic vertex operators

\beq
\left\{ : e^{  \left[ \pm \frac{2}{3} i \vec{\alpha}_j \cdot\vec \psi ( \tau )
\right]} : : e^{ \left[ \mp \frac{2}{3} i \vec{\alpha}_j \cdot
\vec \psi ( \tau' )
\right]} : \right\}_{\tau' \to \tau^-} \approx
\left[ \frac{\pi v ( \tau - \tau' )}{ L } \right]^{- \frac{4g}{9}}
\left[ 1 \pm \frac{2}{3}  ( \tau - \tau' ) \vec{\alpha}_j
\cdot \frac{ \partial \vec{\psi} ( \tau )}{ \partial \tau} \right]
\:\:\:\: .
\label{ju4}
\eneq
\noindent
Terms proportional to $\frac{ \partial \vec{\psi} ( \tau ) }{ \partial \tau}$,
which  could be generated to second-order in $\xi$, are suppressed by the
condition $\sum_{ j =1}^3 \vec{\alpha}_j = 0$. As a result,
higher-order contributions to the $\beta$-function of the running coupling
strength $\zeta = L \xi$ only appears to order $\zeta^3$. The RG equation
for $\zeta$ is then given by

\beq
\frac{ d \zeta }{ d \ell } = [ 1 - h_F ( g ) ] \zeta
- 2 \zeta^3
\:\:\:\: .
\label{ju5}
\eneq
\noindent
For $\gamma = \pi / 3$  the scaling dimension of the boundary interaction,
$h_F ( g)$, gets renormalized as

\beq
\frac{ d h_F (g)}{ d \ell } = -  h_F (g) \zeta^3
\:\:\:\: .
\label{ju6}
\eneq
\noindent
For a small enough value of $\zeta$,  the renormalization of $ h_F (g)$ may
be safely neglected, since it appears only to the third-order in $\zeta$, and
one may substitute $ h_F (g)$ in Eq.(\ref{ju5}) with its bare value
$\frac{4g}{9}$. Thus, the leading boundary perturbation at
the SFP is irrelevant for $g > 9 /4$, while it is
relevant for $g  < 9 / 4$. As a result, for $\gamma = \frac{ \pi}{3}$, there
is a range of values of $g$ -namely, $1 < g < 9/4$-
where neither the WFP, or the SFP,
are stable. The flow diagram then implies the existence of a FFP
 in the phase diagram. In Fig.\ref{jla}, the
phase diagram is sketched for different values of $g$: because of the
periodicity in $\gamma$, only the stripe $ 0 \leq \gamma \leq 2 \pi / 3$ is
drawn.

The new attractive FFP emerges  as a result of the combined
effect of the design of the
YJJN and of the possibility of tuning the frustration parameter
$\gamma$ by setting the dimensionless flux $f$ to $\pi$.
Since the circular array {\bf C} can have a very small diameter,
self-impedance effects should be negligible.
For $Y$-shaped bosonic networks
realized with neutral atomic systems \cite{Demler}, the FFP
is always repulsive, since those systems are  insensitive to external magnetic
fluxes.

\section{The Josephson currents}

In this section, we probe the behavior of a YJJN near each one of its fixed points by computing the current-phase
relation of the Josephson currents along the three arms of  a YJJN.

We find that, for any value of $g$ and for $\gamma \neq
\frac{ \pi}{3}$, the  current-phase relation is the same as the
one of a  Josephson junction chain with a weak link analyzed in
Ref.\cite{giuso1} while,  for $\gamma = \frac{ \pi}{3}$,  one
finds new and unexpected behaviors.

The Josephson currents in the three arms of a $Y$-shaped
JJN are given by

\begin{eqnarray}
I_1 &=& \frac{e^*}{ \beta g} \left[ \frac{ 1}{ \sqrt{2}} \frac{ \partial
\ln {\bf Z}}{ \partial \mu_1} +  \frac{ 1}{ \sqrt{6}} \frac{ \partial
\ln {\bf Z}}{ \partial \mu_2} \right]
\nonumber \\
I_2 &=& \frac{e^*}{ \beta g} \left[ - \frac{ 1}{ \sqrt{2}} \frac{ \partial
\ln {\bf Z}}{ \partial \mu_1} +  \frac{ 1}{ \sqrt{6}} \frac{ \partial
\ln {\bf Z}}{ \partial \mu_2} \right]
\nonumber \\
I_3 &=& - \sqrt{\frac{2}{3} } \frac{e^*}{ \beta g}  \frac{ \partial
\ln {\bf Z}}{ \partial \mu_2}
\:\:\:\: ,
\label{jcw7}
\end{eqnarray}
\noindent
where ${\bf Z}$ is the partition function describing the
thermodynamical behavior of the YJJN,   $\vec{\mu}$ are the
phase differences introduced in section 2.2, and $e^* = 2 e$ is the charge of
a Cooper pair. In the following, we shall provide  the explicit form of
Eq.(\ref{jcw7}) near each one of the three accessible fixed points
analyzed in section 3.

\subsection{The weakly coupled fixed point}

At the WFP, for $g<1$,  $H_B$ is  an irrelevant
perturbation and, thus,

\beq
 {\bf Z} = {\rm Tr} \biggl\{  e^{ - \beta H_0 }   {\bf T}_\tau\exp
\left[ i \int_0^\beta \: d \tau \: H_B ( \tau ) \right] \biggr\}
\:\:\:\: ,
\label{ciro}
\eneq
\noindent
may be safely computed using a  mean-field approximation.
In Eq.(\ref{ciro}),   $H_0 = \frac{ \pi v }{L} \sum_{ j = 1, 2} \:
\sum_{ n = 1}^\infty \alpha_j ( -n + 1 ) \alpha_j ( n )$, and
 the boundary interaction Hamiltonian has been  defined in Eq.(\ref{let3}).
As a result, one gets
\beq
{\bf Z} \approx {\bf Z}_0 \exp \left[ - \int_0^\beta \: d \tau \:
\langle H_B ( \tau ) \rangle_0^{(W)}  \right] =
 {\bf Z}_0  \exp \left[  2 \beta \bar{E}_W \sum_{ i = 1}^3
\cos [ \vec{\alpha}_i \cdot \vec{\mu} + \gamma ] \right]
\:\:\:\: ,
\label{jcw3}
\eneq
\noindent
where $ \langle \ldots \rangle_0$ denotes the thermal average with
Boltzmann weight $e^{ - \beta H_0 }$, and  ${\bf Z}_0 = {\rm Tr}
[ e^{ - \beta H_0 }] = 1 / \{ \prod_{ n = 0}^\infty [ 1 - \bar{q}^{ n +
\frac{1}{2} } ]^2 \}$. From Eqs.(\ref{jcw7},\ref{jcw3}), one gets

\begin{eqnarray}
I_1 &=&  \frac{2 e^* \bar{E}_W }{ g}
\{  \sin [ \vec{\alpha}_1 \cdot \vec{\mu}  + \gamma ]  -
 \sin [ \vec{\alpha}_3 \cdot \vec{\mu}  + \gamma ]  \}
\nonumber \\
I_2 &=&   \frac{2 e^* \bar{E}_W }{ g}
\{  \sin [ \vec{\alpha}_2 \cdot \vec{\mu}  + \gamma ]  -
 \sin [ \vec{\alpha}_1 \cdot \vec{\mu}  + \gamma ]  \}
\nonumber \\
I_3 &=&  \frac{2 e^* \bar{E}_W }{ g}
\{  \sin [ \vec{\alpha}_3 \cdot \vec{\mu}  + \gamma ]  -
 \sin [ \vec{\alpha}_2 \cdot \vec{\mu}  + \gamma ]  \}
\:\:\:\: .
\label{jcw7b}
\end{eqnarray}
\noindent
Eqs.(\ref{jcw7b}) explicitly show the dependence of  the current's patterns
along the arms of the YJJN on both the phase differences $\vec{\mu}$ and
the parameter $\gamma$.

\subsection{The strongly coupled fixed point}

In order to compute the Josephson currents across the three arms of
a YJJN at the SFP, one has now to
account for the contribution coming from the zero modes. In order to do so, one should
use  Eqs.(\ref{jcw7}), with the appropriate expression for
the partition function ${\bf Z}_\ell [ \vec{\mu} ]$ given by
Eq.(\ref{cou4}). The zero modes affect the total energy by the amount

\beq
E_{n_{12} , n_{13} } [ \vec{\mu} ] =
\frac{ \pi v g}{ L} \left[
\left( n_{12} + \frac{\mu_1}{2 \pi} + \frac{ 2 \epsilon_\ell}{3} \right)^2
+ \frac{4}{3} \left( n_{13} + \frac{n_{12}}{2}  +
\frac{\sqrt{3}}{4 \pi} \mu_2 \right)^2 \right]
\:\:\:\: ,
\label{cicillolupo}
\eneq
\noindent
which is a function of $n_{12} , n_{13},  \vec{\mu}$.
At very low temperature and at fixed $\vec{\mu}$, one may
approximate the free energy ($- \frac{1}{ \beta } \ln {\bf Z}$)
 with the lowest value  of the energies $E_{n_{12} , n_{13} }
[ \vec{\mu} ]$, given in Eq.(\ref{cicillolupo}).  For the zero mode
eigenvalues belonging to sublattice A, for instance, the Josephson currents
turn out to be  given by

\begin{eqnarray}
I_1 &=& \frac{e^* v g }{ L } \left[ \frac{1}{ \sqrt{2}}
\left( \frac{\mu_1}{2 \pi }  + n_{12} \right) + \frac{1}{ \sqrt{6}} \left(
\frac{\mu_2}{2 \pi}  + \frac{2 n_{13} + n_{12} }{ \sqrt{3}}
\right)  \right] \nonumber \\
I_2 &=& \frac{e^* v g }{ L } \left[ - \frac{1}{ \sqrt{2}}
\left( \frac{\mu_1}{2 \pi }  + n_{12} \right) + \frac{1}{ \sqrt{6}} \left(
\frac{\mu_2}{2 \pi}  + \frac{2 n_{13} + n_{12} }{ \sqrt{3}}
\right)  \right]
 \nonumber \\
I_3 &=& - \frac{e^* v g }{L  } \sqrt{\frac{2}{3}}  \left(
\frac{\mu_2}{2 \pi}  + \frac{2 n_{13} + n_{12} }{ \sqrt{3}}
\right)
\:\:\:\: .
\label{jsc3}
\end{eqnarray}
\noindent
Eqs.(\ref{jsc3}) show the usual \cite{giuso1} sawtooth dependence on
the phase difference
$\vec{\mu}$, exhibited by the Josephson current at the SFP.
As $\vec{\mu}$ varies within one periodicity interval, the integers
 $n_{12} , n_{13}$ change by $\pm 1$. For instance,
for $-\frac{1}{6} < \frac{ \mu_1}{ 2 \pi} < \frac{1}{6}$,
from $\frac{ \mu_2}{2 \pi}
=  \frac{ \mu_2^*}{2 \pi} - \delta = - \frac{1}{\sqrt{3}} - \delta$ to
$\frac{ \mu_2}{2 \pi} = \frac{ \mu_2^*}{2 \pi} + \delta$ ($\delta / \pi
\ll 1$), the Josephson currents  undergo an abrupt jump from

\beq
I_1 = \frac{ e^* v}{ \sqrt{2} ( 2 \pi ) L } \left( \frac{ \mu_1}{ 2 \pi} -
\frac{1}{3} \right)  \:\:\: , \:\:
I_2 = \frac{ e^* v}{ \sqrt{2} ( 2 \pi ) L } \left( - \frac{ \mu_1}{ 2 \pi} -
\frac{1}{3} \right)
\;\;\; , \;\;
I_3 =  \frac{ \sqrt{2} e^* v}{  6 \pi  L }
\:\:\:\: ,
\label{curdi1}
\eneq
\noindent
to
 \beq
I_1 = \frac{ e^* v}{ \sqrt{2} ( 2 \pi ) L } \left( \frac{ \mu_1}{ 2 \pi} +
\frac{1}{3} \right)  \:\:\: , \:\:
I_2 = \frac{ e^* v}{ \sqrt{2} ( 2 \pi ) L } \left( - \frac{ \mu_1}{ 2 \pi} +
\frac{1}{3} \right)
\;\;\; , \;\;
I_3 = - \frac{ \sqrt{2} e^* v}{  6 \pi  L }
\:\:\:\: ,
\label{curdi2}
\eneq
\noindent
corresponding to the shift $ (n_{12} , n_{13} ) \longrightarrow
 (n_{12} , n_{13} + 1 )$.

It should be noticed that, for $\frac{\mu_1}{2 \pi} = - \frac{1}{3}$,
the current in arm 1 switches from 0 to a finite value, while the current
in arm 2 does the opposite. This suggests that a YJJN may be useful as
a switch commuting  between two states
macroscopically distinguishable by the value of the Josephson current
across the circuit branches. Finally we mention that, as it usually happens in
superconducting networks \cite{giuso2,glhek}, the V-instantons near the
Dirichlet fixed point round off the spikes of the sawtooth function
describing the Josephson current phase relationship. This effect is
discussed in detail in  appendix B.

\begin{figure}
\includegraphics*[width=1.0\linewidth]{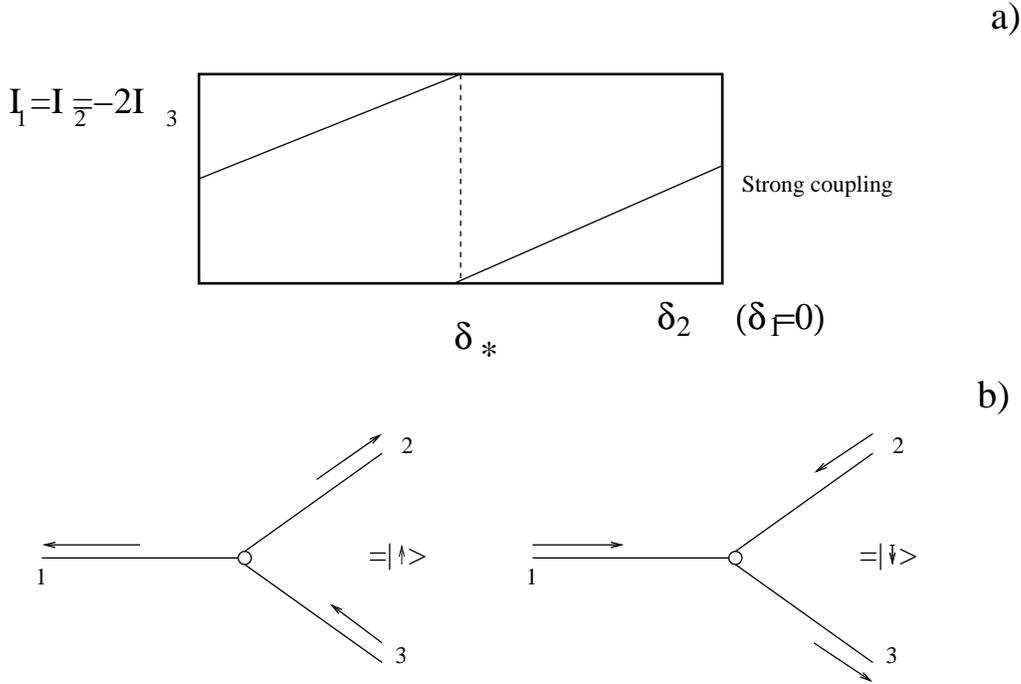}
\caption{{\bf a):} Behavior of the Josephson current across
the three chains near the SFP, with the
phases $\mu_1 ,  \mu_2$ chosen as in section 4.2.
{\bf b):} Sketch of the Josephson current pattern in the YJJN
corresponding to the current patterns across $\mu_2 = \mu_2^*$.}
\label{jllb}
\end{figure}

\subsection{Instanton effects for $\gamma = \pi / 3$
at the finite coupling fixed point}

For $\gamma = \pi / 3$, near the FFP,
new more dramatic instanton effects take place in a YJJN. Indeed,
for $\gamma = \pi / 3$, two triangular sublattices  become degenerate and
shorter instanton paths are allowed.  As evidenced in section 3,
the operators $W_j, W_j^\dagger$ representing
these paths, become relevant for $1 < g < \frac{9}{4}$, and drive the
system away from the Dirichlet point. Here we evidence the
remarkable effects of W-instantons  on the distribution of the
Josephson currents  in  the arms of a YJJN.

For  $\gamma =  \frac{\pi}{ 3}$,  the
minima of the boundary interaction lie on the  honeycomb lattice
depicted in Fig.\ref{ablat}. Their position is given by

\beq
( p_1 , p_2 ) = \sqrt{2g} \left( \left[ n_{12} + \frac{ \mu_1}{ 2 \pi}
+ \frac{ 2 \epsilon}{3} \right] ,  \left[ \frac{ \mu_2}{ 2 \pi}
+ \frac{2}{ \sqrt{3}}
\left(  n_{13} + \frac{n_{12} }{2} \right)  \right]
\right)
\:\:\:\: ,
\label{jsc12}
\eneq
\noindent
with $\epsilon = 0 , 1 $.
Setting $\mu_1 \sim \mu_1^* + \delta =  - \frac{ \pi}{3} - \delta$,
and  $- \frac{ \pi}{ \sqrt{3}} < \mu_2 < \frac{  \pi}{ \sqrt{3}}$,
with $ | \delta | / \pi \ll 1$, the state $ | \uparrow  \rangle =
| 0 , 0 \rangle_A$  and the state $ | \downarrow \rangle 
= |  0 , 0 \rangle_B$  are
quasidegenerate (indeed, they become exactly degenerate for 
$\delta=0$).  For this choice of the quasidegenerate states,
the W-instantons are described by $W ( 0 ) =
- 2 \zeta : \cos \left[ \frac{2}{3} \psi_1 ( 0 ) \right]:$. Substituting
 Eq.(\ref{cou48}) into Eq.(\ref{c7}) allows to write their contribution
to the partition function as

\begin{eqnarray}
{\bf Z} [ \delta , \mu_2 ]  &=& \sum_\sigma
\int_{ - \infty}^\infty \: d x \:  \frac{
e^{ i \frac{g  \pi v \beta}{2 L} x } }{ 2 \pi}  \biggl\{  i x + e_0
 + {\rm sg} ( \sigma ) \alpha \delta - \zeta^2 \Gamma [ 1 - 2 
h_F ( g )] [ i x + e_0  +
{\rm sg} ( \sigma ) \alpha \delta ]^{ 2 h_F ( g )
- 1} \biggr\}  \biggl/ \nonumber \\  &~& \biggl\{
[ i x + e_0 ]^2 - [ \alpha^2 \delta^2
+ \zeta^2 ] - \zeta^2 \Gamma [ 1 - 2 h_F (g) ] \sum_{ \gamma }
[ i x + e_0  + {\rm sg} ( \gamma )  \alpha \delta
]^{2 h_F ( g ) } \nonumber \\
  &-& \zeta^4  \Gamma^2 [ 1 - 2 h_F (g) ]
[ ( i x + e_0 )^2 - \alpha^2 \delta^2 ]^{ 2 h_F ( g ) -1}
\biggr\}
\:\:\:\: ,
\label{cou51}
\end{eqnarray}
\noindent
where  $\alpha = g / ( 6 \pi )$,
$e_0 \equiv e_0 ( \delta , \mu_2 ) = g  \left( \frac{1}{9} +
\frac{\delta^2}{ 4 \pi^2} +\frac{\mu_2^2}{ 4 \pi^2} \right)$,
 $x =  \frac{ 2 L}{ \pi v }  \omega$, and 
$ {\rm sg} ( \sigma ) = 1$ if $\sigma = \uparrow$, $=-1$, if $\sigma
= \downarrow$.

\begin{figure}
\includegraphics*[width=1.0\linewidth]{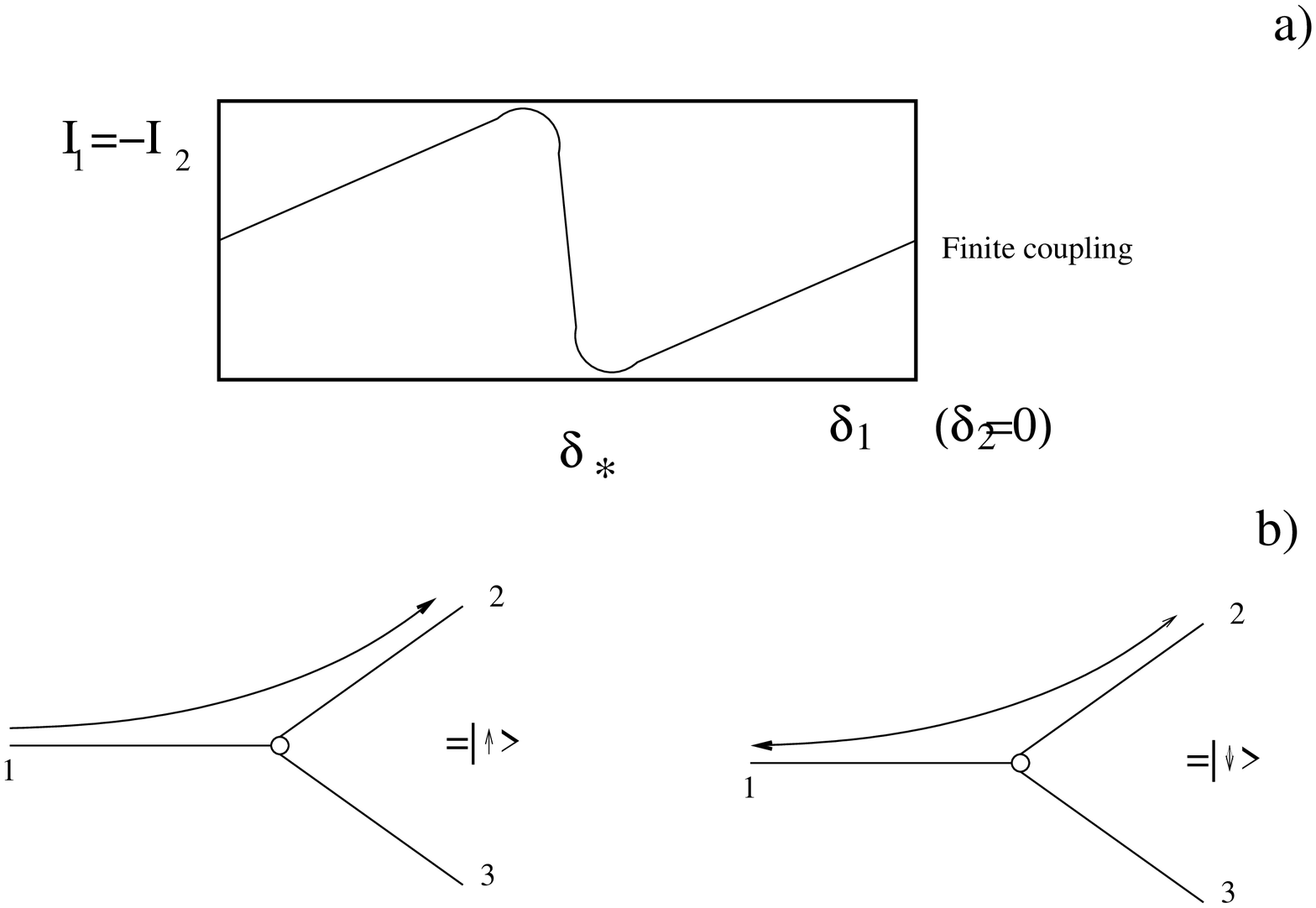}
\caption{{\bf a):} Behavior of the Josephson current across
the chains 1 and 2 near the FFP, with the
phases $\mu_1 ,  \mu_2$ chosen as discussed in section 4.3.
{\bf b):} Sketch of the Josephson current pattern in the YJJN
associated to the macroscopic
states $ | \sigma ( = \uparrow , \downarrow ) \rangle$
introduced in section 4.3.}
\label{jlla}
\end{figure}

To compute Eq.(\ref{cou51}) is quite a formidable task:
however, an approximate  computation can be carried out, for
$g=\frac{9}{4} - \epsilon$ ($ \epsilon \ll 1$),
near the FFP $\zeta_*
= ( 1 -\frac{4g}{9})^\frac{1}{2}$. Indeed,
since = $\zeta_* \sim \epsilon^\frac{1}{2}  \ll 1$,
neglecting  $O ( \zeta_*^4)$-terms in Eq.(\ref{cou51}), leads to

\beq
{\bf Z} [ \delta , \mu_2  ] \approx
 \exp \left[ - \beta \frac{ \pi v }{L} e_0 ( \delta , \mu_2  )
\right] \left\{ \cosh \left[ \beta \frac{ \pi v \alpha}{ L } \delta \right]
+ \cosh \left[ \beta \frac{ \pi v \alpha}{ L } \sqrt{ \alpha^2 \delta^2 +
3 \zeta_*^2 } \right] \right\}
\:\:\:\: ,
\label{jsc23}
\eneq
\noindent
from which, for $\beta v / L \gg 1$, one gets

\begin{eqnarray}
I_1 & \approx & \frac{e^* v}{ 2 \pi L } \left\{ \frac{\delta}{ \sqrt{2}}
\left[ -1 + \frac{ 4 \pi^2 \alpha^2}{ \sqrt{ \alpha^2 \delta^2 + 3
\zeta_*^2} } \right] -  \frac{\mu_2}{ \sqrt{6}} \right\}  \nonumber \\
I_2 & \approx & \frac{e^* v}{ 2 \pi L } \left\{ - \frac{\delta}{ \sqrt{2}}
\left[ -1 + \frac{ 4 \pi^2 \alpha^2}{ \sqrt{ \alpha^2 \delta^2 + 3
\zeta_*^2} } \right] -  \frac{\mu_2}{ \sqrt{6}} \right\} \nonumber \\
I_3 & \approx & \frac{e^* v}{ 2 \pi L } \sqrt{\frac{2}{3}} \mu_2
\:\:\:\: .
\label{jsc24}
\end{eqnarray}
\noindent
Eqs.(\ref{jsc24}) yield the current-phase relations near the attractive
FFP shown in Fig.\ref{jlla}. The typical sawtooth behavior
of the Josephson current-phase relation is now associated to a stable
attractive FFP in the phase diagram.

Usually,  in superconducting systems, such as SQUIDs \cite{glhek,giuso2} and
Josephson chains with localized  impurities \cite{glark,giuso1}, either the
smoothening of the spikes of the sawtooth function describing the  Josephson
current-phase relationship at strong coupling
is a perturbative effect, or the SFP is unstable, since quantum
fluctuations drive the system to the WFP. At variance,
for a YJJN at the FFP, the smoothening of the spikes of
the Josephson current due (now) to the W-instantons is
a nonperturbative effect and the FFP is a stable attractive
fixed point in the phase diagram.  Since a
sawtooth behavior of the Josephson current is usually
associated \cite{glhek,giuso1,giuso2} to the emergence of a macroscopically
quantum coherent two-level system in the superconducting device
\cite{shon0}, one may safely expect that an effective macroscopic
two-level quantum system -this time robust against quantum 
fluctuations- may emerge  in a YJJN, as well.  This issue
has been addressed in Ref.\cite{giusonew} and will be revisited in 
the next section.

\section{A quantum two-level system emerging in a YJJN}

In this section, we revisit the arguments given in 
Ref.\cite{giusonew} to show that an effective quantum two-level
system with frustrated decoherence \cite{novais} may emerge from a YJJN 
near the FFP.

As evidenced in appendix A, the energy of the long-wavelength excitations
of the YJJN is given by

\beq
E = \frac{ \pi v }{ 2 L } [ \vec{p} ]^2 + E'
\;\;\;\; ,
\label{formaggio}
\eneq
\noindent
 where $\vec{p} = (p_1 , p_2 )$ is the eigenvalue of the zero-mode operators
$\vec{P} = ( P_1 , P_2 )$, introduced in section 3,
and $E'$ accounts for  the energy of the plasmon modes described by the TLL Hamiltonian given by Eq.(\ref{adi54.b}).
 Plugging Eqs.(\ref{subla},\ref{sublb}) into Eq.(\ref{formaggio})
yields the explicit dependence of the energy on the minima on
the phases $\varphi_j$ of the three bulk superconductors.

It is easy to see that, for any value of $\gamma$ and for all possible
values of the Luttinger parameter $g$, it is always possible,
for a finite YJJN, to choose the phase differences $\mu_1$ and $\mu_2$
to obtain two low-energy quasidegenerate states, well separated
from the rest of the spectrum. Slightly generalizing the notation introduced in Section 4, we still denote
the two quasidegenerate states by $ | \uparrow \rangle$ and $ | \downarrow
\rangle$ and observe that, for $\gamma \neq ( 2 k + 1 ) \pi / 3$,
both states belong to the same triangular sublattice (A, B or C),
while, for  $\gamma = ( 2 k + 1 ) \pi / 3$, they belong - as in Section 4- to two
different sublattices.

The dynamics of the two states $ | \uparrow \rangle$ and $ | \downarrow
\rangle$ interacting with the plasmon modes residing on the three chains of the Y-junction
may be written as

\beq
H_2 = [ E_\uparrow ( \vec{\mu} ) -  E_\downarrow ( \vec{\mu} ) ]
\sigma^z - Y {\bf O} ( 0 ) \sigma^- -  Y {\bf O}^\dagger ( 0 ) \sigma^+
\:\:\:\: ,
\label{c1}
\eneq
\noindent
where
 ${\bf O}$ is one of the vertex operators $V_j$ (if the states
lie on the same triangular sublattice), or  $W_j$ (if the states
lie on the honeycomb lattice obtained by merging  two
triangular sublattices), and $\sigma^z = \sum_{\sigma = \uparrow, \downarrow}
\frac{1}{2}  {\rm sg} ( \sigma ) | \sigma \rangle \langle \sigma |$,
$ \sigma^+ = | \uparrow \rangle \langle \downarrow |$,
$ \sigma^- = | \downarrow \rangle \langle \uparrow |$.
 In Eq.(\ref{c1}), $E_\uparrow
(  \vec{\mu} ) $ and  $E_\downarrow
(  \vec{\mu} ) $ are the energies associated to the
states $ | \uparrow \rangle$ and $ | \downarrow \rangle$,
respectively. The $\sigma^z$-term contributes only
if the $ | \uparrow \rangle$ and the $ | \downarrow \rangle$ states
are quasidegenerate, which may be achieved by a slight detuning of
the phase differences $\mu_1$ and $\mu_2$ by an amount
$\delta$ ($ \ll 2 \pi $). The terms proportional to $Y$ describe
an effective field in the $x$-direction and, at the
same time, the coupling between the transverse components of the spin
and the bath provided by the plasmon modes of the three chains: on one hand they determine a
$Y$-dependent renormalization of the energies of the effective
two-level system -the tunnel splitting of the energies of the
states $ | \sigma \rangle$-, on the other hand, they may lead to
the formation of an entangled  state between the two-level system and
the bath formed by the plasmon modes in the network. This latter effect is a main source of
decoherence in a two-level system interacting with one (or more) baths
\cite{novais,giusonew}.

Depending on whether $\gamma \neq ( 2 k + 1 ) \pi /3$, or $\gamma
= ( 2 k + 1 ) \pi / 3$ and on the value of the Luttinger parameter $g$
 the interaction of the system with the
bath provided by the plasmon modes of the network leads to  different coherent behaviors of the YJJN \cite{giusonew}.
In the following we shall compute the spectral density of the two level system near the SFP and the FFP; as pointed out
in Ref.\cite{novais},
the spectral density provides a measure of the amount of entanglement between a two level system and the
pertinent environmental modes.

\subsection{Spectral density of the two-level system near the
strongly-coupled fixed point}

As evidenced in section 3,  for $1 < g < 9 /4$ and
$\gamma \neq ( 2 k + 1 ) \pi / 3$, or for $g > 9 /4$
and $\forall \gamma$,   the YJJN exhibits
an IR  stable SFP in its phase diagram. If $ - \pi / 3 < \gamma < \pi /3$,
Eq.(\ref{subla}) implies that the quasidegenerate states
$ | \uparrow \rangle$ and $ | \downarrow \rangle$ lie on
the triangular sublattice A and that Eq.(\ref{c1}) may be explicitly
written as

\beq
H_{2 , {\rm SFP}} =
[ E_\uparrow ( \vec{\mu} ) -  E_\downarrow ( \vec{\mu} ) ]
\sigma^z - Y V ( 0 ) \sigma^- -  Y V^\dagger ( 0 ) \sigma^+
\:\:\:\: ,
\label{cici1}
\eneq
\noindent
where $V ( 0 )$ is the V-instanton vertex operator.

To compute the spectral density
of the two-level system in Eq.(\ref{cici1}) one needs to evaluate
$\chi^{``}_\perp ( \Omega ) / \Omega $ {\it vs.} $\Omega$, where
$\chi^{``}_\perp ( \Omega ) $ is the imaginary part of the transverse
dynamical  spin susceptibility. Since V-instantons
are an irrelevant perturbation, by neglecting higher-order corrections
in $Y$ (see appendix B), one gets

\beq
\frac{ \chi^{``}_\perp ( \Omega ) }{ \Omega}
  \propto \delta ( \Omega - 2 \Delta ( \vec{\mu} ) )
+ \delta ( \Omega + 2 \Delta ( \vec{\beta } ) )
\:\:\:\: ,
\label{dd3}
\eneq
\noindent
with  $\Delta ( \vec{\mu} ) = \sqrt{ [ E_\uparrow ( \vec{\mu} ) ]^2
+ Y^2 }$.
From Eq.(\ref{dd3}) one sees that the spectrum of Eq.(\ref{cici1})
 is given by two
classical  states, with $\Omega = \pm \Delta ( \vec{\mu})$. 
As pointed out in ref. \cite{giusonew}
this behavior signals that there is no entanglement between the 
two level quantum system and the plasmon modes. Since $Y$ is irrelevant 
(i.e., its fixed point value is $Y_* = 0$),
 there is not even tunnel splitting between the
 two degenerate states, no quantum coherence may emerge
in this regime \cite{giusonew}

\subsection{Spectral density of the two-level system near the
finite-coupling fixed point}

\begin{figure}
\includegraphics*[width=1.0\linewidth]{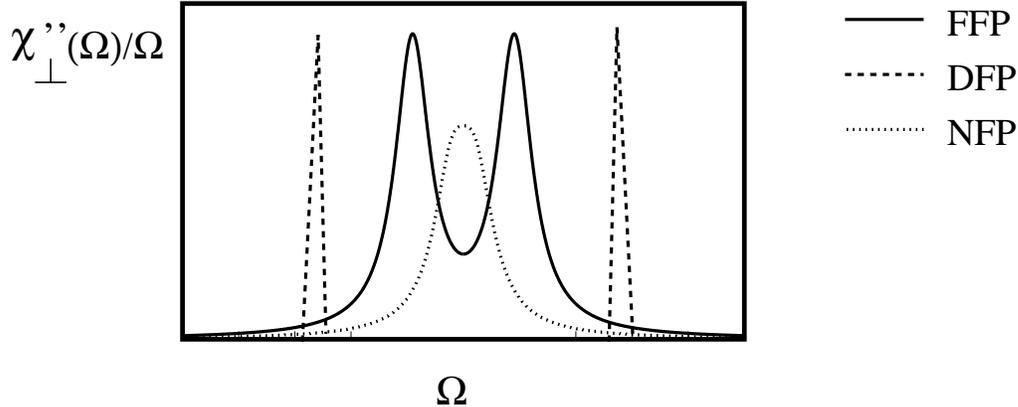}
\caption{Behavior of the spectral density
$\chi^{``}_\perp ( \Omega ) / \Omega$ vs. $\Omega$
near the WFP, the SFP, and the FFP.}
\label{deco}
\end{figure}

For $1 < g < 9 / 4$ and $\gamma = \pi / 3$, the two states
$ | \uparrow \rangle$ and $ | \downarrow \rangle$
lie on nearest neighboring sites on the
honeycomb lattice obtained by merging the sublattices A and B.
As evidenced in section 3, short W-instantons are a relevant
perturbation at the SFP and render the FFP IR stable. Near the
FFP, the two-level system is described by

\beq
H_{2 , {\rm FFP}} =
[ E_\uparrow ( \vec{\mu} ) -  E_\downarrow ( \vec{\mu} ) ]
\sigma^z - \xi W ( 0 ) \sigma^- -  \xi W^\dagger ( 0 ) \sigma^+
\:\:\:\: ,
\label{cici2}
\eneq
\noindent
where, now, $W ( 0 )$ is a W-instanton operator. The
computation of the spectral density  $\chi_\perp^{``} ( \Omega )$
is detailed in appendix B using  a self-consistent RPA approximation.
As a result, the spectral density has now two peaks centered at
 $\pm \Delta_*  ( \vec{\mu} ) = \pm \sqrt{ [ E_\uparrow (
\vec{\mu} ) ]^2 + ( \zeta_* / L )^2}$, with a finite width
$\propto \frac{ \pi v}{L}
( \zeta_* )^{ 1 + \frac{8}{9} g}$, where $\zeta_*$ is the finite fixed
point value of the running coupling constant, determined in section
3. The spectral density is plotted in Fig.\ref{deco}, where
we report, for completeness, also the spectral density arising near the
WFP \cite{giusonew}.

\section{Concluding remarks}

We showed that, for $1<g< 9 / 4$ and for $f = \pi$, an attractive
FFP emerges in the phase diagram accessible to a YJJN. The
new fixed point does not break time-reversal invariance, and it
is a stable, attractive fixed point only when the dimensionless flux threading the cental loop equals $\pi$; for
$f \neq \pi$, we have shown that only the SFP survives as a stable attractive fixed point.
Our results show remarkable similarities between the phase diagram
accessible to a YJJN and the one arising in the analysis of the
quantum Brownian motion  on a frustrated triangular lattice
\cite{kaneyi,saleurb}.

Crucial to our analysis is the fact that, at $f = \pi$ - for
$1<g< 9 / 4$ - the W-instantons become a relevant perturbation and, thus, destabilize the SFP.
These instantons emerge ultimately as a result of the  Y-shaped geometry of the network,
since they arise when the minima of the boundary potential
span the honeycomb lattice depicted in Fig.\ref{ablat}. Intuitively,
they may be regarded as the result of the ``deconfinement'' of the
V-instantons in its elementary constituents happening- when $1<g< 9 / 4$- only at $f = \pi$. A Coulomb gas
approach could be a very helpful tool to further clarify the nature of the FFP.

We computed the current-phase relations along the arms of the YJJN near
each one of the allowed fixed points. We evidenced the parameter
regions where a YJJN may be operated as a Josephson switch and we
showed the different effects of the instantons on the current pattern
near the SFP and the FFP. In particular, in  a YJJN at the FFP, the
smoothening of the spikes of the sawtooth dependence of Josephson current on the phase differences $\vec{\mu}$
is a nonperturbative effect, due to the attractive nature of this fixed point.

Finally, we provided additional arguments confirming that, near the FFP, a YJJN supports a quantum coherent
two-level system with frustrated decoherence \cite{giusonew}.

In order to set a YJJN to be a quantum device either acting as a Josephson current switch or modeling an effective two level
 quantum system
one needs, first of all, to promote the phase differences $\mu_1 , \mu_2$ to control parameters. This may be
achieved by resorting, for instance, to multipolar magnetic coils 
\cite{granata} inserted in external loops connecting the
bulk superconductors at the outer boundary of the YJJN: indeed, for
sufficiently long chains, the localized magnetic fields generated by the
 multipolar magnetic coil
may be engineered to avoid variations in the flux threading the circular 
Josephson junction
array {\bf C}. Furthermore, when the YJJN has a finite size $L$, it 
is easy to convince oneself that the FFP is stable
against small fluctuations of the flux $f$, provided
that $v /L$ is sufficiently big: for instance, if the point
 $\gamma = \pi / 3$ is
displaced by a small amount $\nu$, $v / L $ needs to be larger than
the energy splitting  $\bar{E}_W \sin ( \nu )$ between the minima of two
triangular sublattices. At variance, when $v/L < \bar{E}_W \sin (
\nu)$, there is a flow towards the SFP and,  depending on
${\rm sgn} (\nu)$, the minima of the boundary potential lie on
either one of the triangular A and B  sublattices \cite{giusonew}. 
Finally, today 's technology allows to fabricate
superconducting devices with values of $g$ ranging from $g < 1$,
 to $g \sim 2$ \cite{haviland}.

Josephson networks where $n$ finite chains are connected to a
central circular array ${\bf C}$ may be analyzed with tools similar to those used in this paper. Of interest
is also the JJ network with $n=4$ since it corresponds to the tetrahedral qubit proposed in Ref.\cite{blatter}.

\label{}

\vspace{1cm}

{\bf Acknowledgments }:
 We thank I. Affleck, C. Chamon, R. Russo  and
A. Trombettoni for  fruitful discussions and correspondence. 
We thank the Particle Theory Sector of S.I.S.S.A. - I.S.A.S.
and the High Energy Theory Group of I.C.T.P. for hospitality 
during the final stages of our work.

\vspace{0.5cm}

 \appendix

\section{Tomonaga-Luttinger description of superconducting
Josephson junction arrays}

Here, we briefly review the derivation of the effective TLL 
Hamiltonian, describing  one-dimensional arrays of
Josephson junctions. For this purpose,  in Eq.(\ref{eql.1}) one
should assume that  $E_J / E_c \ll 1$ and ${\bf N} = 2 n + 1 + 2 h$,
 with $n$ integer and $ | h | \ll 1$  \cite{glark,giuso1};
then, if one defines effective lattice
spin-1/2 operators as

\beq
  S_j^z = {\bf P}_G \left[ - i \frac{ \partial }{ \partial \phi_j }
- \frac{{\bf N}}{2} \right] {\bf P}_G \;\;\; , \;\;
 S_j^\pm = {\bf P}_G e^{ \pm i \phi_j } {\bf P}_G \;\;\;\; ,
\label{aa.1}
\eneq
\noindent
with ${\bf P}_G$ the operator projecting onto the subspace of the
charge eigenstates with the charge at any
site either equal to $n$ or to $n+1$, one may present Eq.(\ref{eql.1}) as

\[
 P_{\bf G}^\dagger H_{\rm chain}  P_{\bf G} \equiv H_{\rm spin} =
- \frac{E_J}{2} \sum_{ j = 1}^{ L / a - 1 }  [ S_j^+ S_{j+1}^-
+ S_{ j + 1 }^+  S_j^- ]
\]

\beq
+ \left[ E_z - \frac{3}{16} \frac{(E_J)^2}{E_c} \right]
 \sum_{ j = 1}^{ L / a - 1 }  S_j^z S_{j+1}^z
- H  \sum_{ j = 1}^{ L / a }  S_j^z
\:\:\:\: .
\label{eql.4}
\eneq
\noindent
Eq.(\ref{eql.4}) is  the Hamiltonian for an XXZ-chain in an external magnetic
field $H = h E_c$  \cite{glark,giuso1}: to map it  onto an effective TLL
Hamiltonian, one needs to write the spin operators in terms of
lattice Jordan-Wigner fermions $a_j$. Upon defining  the  lattice Fourier
modes $a_k$ as

\beq
a_k = \sqrt{ \frac{a}{L}} \sum_{ j = 1}^{L/a} a_j e^{ - i k ( j a ) }
\:\:\: ; \:\: ( k = \frac{ 2 \pi n}{L} \:  , \; n = 1 , \ldots , L/a )
\:\:\:\: ,
\label{eql.7}
\eneq
\noindent
$  P_{\bf G}^\dagger H_{\rm chain}  P_{\bf G} $ in Eq.(\ref{eql.4}) is given by
\beq
H_{\rm JW} = \sum_k [ - E_J \cos ( k a ) - H ] a^\dagger_k a_k
+ \left[ E_z - \frac{3}{16} \frac{(E_J)^2}{E_c} \right]
 \sum_{ j = 1}^{ L / a - 1 } ( a_j^\dagger a_j - \frac{1}{2} )
 ( a_{j+1}^\dagger a_{j+1} - \frac{1}{2} )
\:\:\:\: .
\label{eql.8}
\eneq
\noindent
From Eq.(\ref{eql.8}), one see that two "band-insulating"
phases open up when $|H| \geq E_J$ \cite{glark,giuso1}.
In spin coordinates, they correspond to fully polarized
spin phases which are  the Coulomb blockade insulating phases
setting in the chain when the gate voltage is tuned far from
charge degeneracy point.

For  $|H| < E_J$, Eq.(\ref{eql.8}) describes  a one-dimensional
conductor. By keeping only long-wavelength modes around the Fermi points
$k_f^\pm = \pm \frac{1}{a} {\rm arccos} ( H / E_J)$, and by
bosonizing Eq.(\ref{eql.8}) one gets  the Sine-Gordon Hamiltonian

\beq
H^b = \frac{g}{ 4 \pi } \: \int_0^L \: d x \:
\left[  \frac{1}{v}\left( \frac{ \partial \Phi }{ \partial t} \right)^2 +
v \left( \frac{ \partial \Phi }{ \partial x} \right)^2  \right]
- G_U \: \int_0^L \: d x \: \cos [ 2 \sqrt{2} g \Phi ( x ) + 4 k_f x]
\:\:\:\: ,
\label{a01.1}
\eneq
\noindent
with the Luttinger parameter defined in section 2 and $G_U
\propto \left[ E^z -
\frac{3}{16} \frac{(E_J)^2}{E_c} \right]$. When  $g>1/2$,
the last term in Eq.(\ref{a01.1}) may  be neglected,
in the thermodynamic limit and   $H^b$ reduces to  the
 Hamiltonian of a spinless TLL.
$g$ may either be $<1$, or $>1$, depending on whether $\Delta ( = \left[ E^z -
\frac{3}{16} \frac{ (E_J)^2}{E_c} \right]) > 0 $ (repulsive TLL),
or $\Delta < 0$ (attractive TLL) \cite{giuso1}.

The normal modes of a spinless TLL may be
constructed by introducing
the dual field $\psi$, related to $\Phi$ by $\frac{1}{v} \frac{ \partial
\psi}{ \partial t} =  \frac{ \partial  \Phi}{ \partial x}$ and
  $\frac{1}{v} \frac{ \partial \Phi}{ \partial t} =
\frac{ \partial  \psi}{ \partial x}$, and by introducing
two chiral bosonic fields, $\phi_R , \phi_L$, as
\beq
\phi_R ( x ) = \sqrt{ \frac{g}{2}} \Phi ( x ) + \frac{1}{ \sqrt{2g}}
\psi ( x ) \;\;\; ; \;\;
\phi_L ( x ) = \sqrt{ \frac{g}{2}} \Phi ( x ) - \frac{1}{ \sqrt{2g}}
\psi ( x )
\;\;\;\; .
\label{eql.14}
\eneq
\noindent
In terms of $\phi_R , \phi_L$, $H^b$ is given by

\beq
H^b = \frac{v}{ 4 \pi } \: \int_0^L \: d x \: \left[ \left(
\frac{ \partial \phi_R  }{ \partial x} \right)^2 + \left(
\frac{ \partial \phi_L  }{ \partial x} \right)^2  \right]
\:\:\:\: .
\label{eql.15}
\eneq
\noindent
The normal mode  expansion of $\phi_R ( x - v t ) ,
\phi_L ( x + v t ) $ may be written in terms of the
Fubini-Veneziano chiral fields \cite{fubini} as

\[
\phi_R ( x - v t ) = q_R - \frac{ 2 \pi}{L} P_R ( x - v t )
+ i \sum_{ n \neq 0} \frac{ \alpha_R ( n )}{ n } e^{ i k_n ( x - v t )}
\]
\noindent
\beq
\phi_L ( x + v t ) = q_R + \frac{ 2 \pi}{L} P_L ( x + v t )
+ i \sum_{ n \neq 0} \frac{ \alpha_L ( n )}{ n } e^{ i k_n ( x + v t )}
\:\:\:\: ,
\label{eql.16}
\eneq
\noindent
with
\beq
[ q_R , P_R ] = [ q_L , P_L ] = i \;\;\; ; \;\;
[ \alpha_R ( n ) , \alpha_R ( m ) ] = - [ \alpha_L ( n ) , \alpha_L ( m ) ]
= n \delta_{ n + m , 0 }
\;\;\;\; ,
\label{eql.17}
\eneq
\noindent
with all the other commutators vanishing.
As a result:

\beq
H^b = \frac{\pi v}{L} [ (P_R)^2 + (P_L )^2 ] + \frac{ \pi v }{L}
\sum_{ n \neq 0 } [ \alpha_R ( - n ) \alpha_R ( n ) + \alpha_L ( n )
\alpha_L ( - n ) ]
\:\:\:\: .
\label{eql.17.b}
\eneq
\noindent
To construct the Fock space, one needs to define a
vacuum  $ | ( p_R , p_L) , 0 \rangle$  for any allowed pair of eigenvalues
of the zero-mode operators $P_R, P_L$, and then act with creation
operators $\alpha_R ( n ) , \alpha_L ( - n )$ ($n<0$) on
the states   $ | ( p_R , p_L) , 0 \rangle$, which obey the
conditions

\[
P_R |  ( p_R , p_L) , 0 \rangle = p_R  |  ( p_R , p_L) , 0 \rangle \;\;\; ,
\;\; P_L |  ( p_R , p_L) , 0 \rangle = p_L  |  ( p_R , p_L) , 0 \rangle
\:\:\:\;  ,
\]
\beq
\alpha_R ( n )  |( p_R , p_L)  , 0 \rangle =
 \alpha_L ( - n )  | ( p_R , p_L) ,  0 \rangle = 0 \;\;\; ( n  > 0 )
\;\;\;\; ,
\label{eql.18}
\eneq
\noindent

In a system with boundaries,
the boundaries conditions may be accounted for by means of pertinent
relations between the $_R$ and the $_L$ modes. For instance, Neumann
boundary conditions at $x=0$, that is, $\frac{ \partial \Phi ( 0 ) }{ \partial
x} = 0$, imply

\beq
P_R - P_L = 0 \;\;\; ,\;\; \alpha_R ( n ) + \alpha_L ( - n ) = 0 \;\; , \;
\forall n
\;\;\;\; ,
\label{neum}
\eneq
\noindent
while Dirichlet boundary conditions  at $x=0$, that is, $\Phi ( 0 ) = 0$,
imply

\beq
P_R + P_L = 0 \;\;\; ,\;\; \alpha_R ( n ) - \alpha_L ( - n ) = 0 \;\; , \;
\forall n
\;\;\;\; .
\label{diri}
\eneq
\noindent

\section{The partition function and the spectral density of  
the effective two-state system}

Here we  set up the  general formalism needed to include the
instanton contributions to the partition function of the effective
two-level system described in section 5. In doing so,
it is most convenient to write  the spin-1/2
operators introduced in Eq.(\ref{c1}) by means of two pairs of
fermionic operators, $ a_\sigma , a^\dagger_\sigma$, such that
$\sigma^z = \frac{1}{2} \sum_{ \sigma = \uparrow \downarrow} {\rm sg} ( \sigma)
 a_\sigma^\dagger a_\sigma$,
$\sigma^+ = a_\uparrow^\dagger a_\downarrow$. When doing so, the
imaginary time action of the effective two-level system reads as

\beq
S_E = \int_0^\beta \: d \tau \: \left\{  \sum_\sigma \: a_\sigma^\dagger
\left[ \frac{ \partial }{ \partial \tau} - i \omega_0 -
E_\sigma ( \vec{\mu}) \right] a_\sigma + [ a_\uparrow^\dagger a_\downarrow
{\bf O} ( \tau ) + {\rm h.c.} ] \right\} + S_E^{(0)}
\:\:\:\: ,
\label{c2}
\eneq
\noindent
where $ S_E^{(0)} $ is the Euclidean action for the plasmon field, given
by

\beq
 S_E^{(0)} =  \sum_{ j =1, 2, 3} \frac{g}{ 4 \pi} \: \int_0^\beta \:
d \tau \:  \: \int_0^L \: d x \: \left[
\frac{1}{v} \left( \frac{ \partial \Phi_j }{ \partial \tau} \right)^2
+ v  \left( \frac{ \partial \Phi_j }{ \partial x} \right)^2 \right]
\;\;\;\; ,
\label{adi54.bap}
\eneq
\noindent
while the chemical potential is $i \omega_0 = i \frac{ \pi}{ \beta}$
\cite{novais}.

At low temperature ($\beta \frac{v}{L} \gg 1$), one  may
approximate the partition function of  the effective
two-level system as

\beq
{\bf Z}_{\rm Eff} \approx \sum_\sigma {\bf Z}_{ \sigma \sigma}
 ( \beta ) \;\;\;\; ,
\label{c3}
\eneq
\noindent
with ${\bf Z}_{ \sigma \sigma'} ( \tau ) = \langle a_\sigma ( \tau )
a_{\sigma'}^\dagger ( 0 ) \rangle$.

\begin{figure}
\includegraphics*[width=1.0\linewidth]{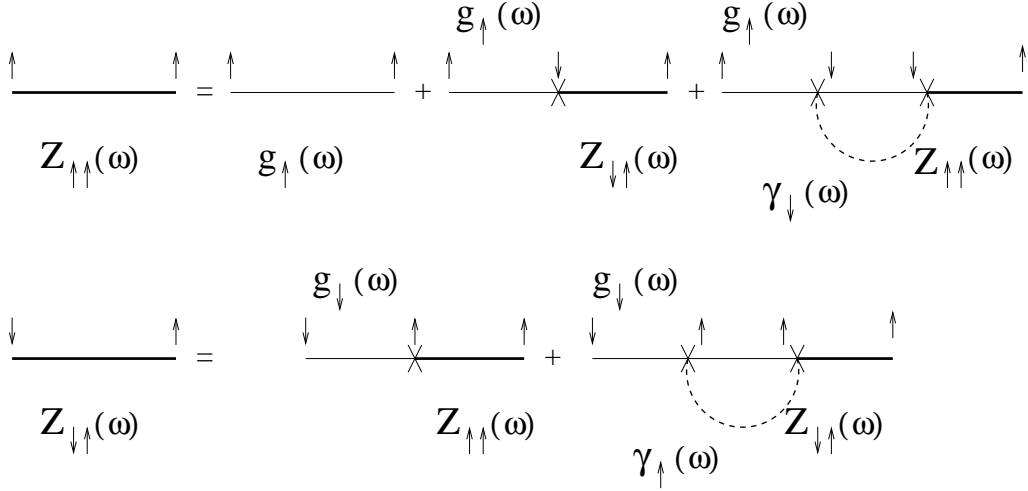}
\caption{Diagrams contributing to  the Dyson equations for
${\bf Z}_{\uparrow \uparrow}$ and for ${\bf Z}_{\downarrow \uparrow}$.}
\label{dyson}
\end{figure}

The  diagrams used to compute
${\bf Z}_{\sigma \sigma'}  ( \omega ) = \int_0^\infty \: d \tau \:
e^{ - i \omega \tau } {\bf Z}_{ \sigma \sigma'} ( \tau )$ are
schematically depicted in
Fig.\ref{dyson}; there, the solid thin line corresponds to the propagator
of  a fermion $a_\sigma$, given by $g^{(0)}_\sigma ( \omega ) =
1 / [  i ( \omega - \omega_0 ) + E_\sigma ( \vec{\mu} ) ]$, and
the dashed line corresponds to the propagator for the $Y$-vertex. The
pertinent Dyson's  equations yielding ${\bf Z}_{ \sigma \sigma'} ( \omega )$
are given by

\begin{eqnarray}
{\bf Z}_{ \sigma \sigma} ( \omega ) &=&
g_\sigma^{(0)} ( \omega ) \left\{ 1 + Y
{\bf Z}_{ \bar{\sigma} \sigma} ( \omega ) + Y^2
\gamma_{ \bar{\sigma}} ( \omega )  {\bf Z}_{ \sigma \sigma} ( \omega )
\right\} \nonumber \\
{\bf Z}_{ \bar{\sigma} \sigma} ( \omega ) &=&
g_{\bar{\sigma}}^{(0)} ( \omega ) \left\{  Y
{\bf Z}_{ \sigma \sigma} ( \omega ) + Y^2
\gamma_{\sigma} ( \omega )  {\bf Z}_{ \bar{\sigma} \sigma} ( \omega )
\right\}
\;\;\;\; ,
\label{c4}
\end{eqnarray}
\noindent
where $\bar{\sigma}$ is $\uparrow ( \downarrow) $ if $\sigma$ is $
\downarrow ( \uparrow)$,  $\gamma_\sigma ( \omega )$ corresponds to the
``bubble'' diagram in Fig.\ref{dyson}  given by

\[
\gamma_\sigma ( \omega ) = \int_{ 0 }^\infty \: d \tau \: \theta ( \tau )
\: e^{ - i \omega \tau } \: \gamma_\sigma ( \tau ) \]
\beq
 =
\frac{ 2 L}{ \pi v} \: \left[
\frac{ \Gamma [ 1 - 2 h (g) ]}{ \frac{2 L}{\pi v}
( i ( \omega - \omega_0) + E_\sigma ( \vec{\mu})  ) - h ( g )  }
\right] \left[ \frac{ \Gamma [ \frac{2 L}{ \pi v } ( i ( \omega - \omega_0)
 + E_\sigma ( \vec{\mu})  )
+ h ( g )  ] }{ \Gamma [ \frac{2 L}{ \pi v } ( i ( \omega - \omega_0)
+ E_\sigma  ( \vec{\mu} ) - h ( g ) ] } \right]
\:\:\:\: ,
\label{cou47}
\eneq
\noindent
with $ h ( g ) = h_S ( g ) = \frac{4}{3} g$ for V-instantons, while
$h ( g ) = h_F ( g ) = \frac{9}{4} g$ for W-instantons. When
 $L \omega / v \gg 1$, Eq.(\ref{cou47})  may be approximated as

\beq
\gamma_\sigma ( \omega ) \approx \left( \frac{2 L}{ \pi v }
\right)^{2 h ( g ) } \:  \Gamma [ 1 - 2 h (g) ] \:
[ i ( \omega - \omega_0) + E_\sigma ( \vec{\mu} )  ]^{ 2 h
( g ) - 1 }
\:\:\:\: .
\label{cou48}
\eneq
\noindent
From Eqs.(\ref{c4}), one obtains

\[
{\bf Z}_{\sigma \sigma} ( \omega ) = \{ [ g_{\bar{\sigma}}^{(0)}]^{-1}
( \omega ) - Y^2 \gamma_\sigma ( \omega ) \} \biggl/
\]
\beq
\left\{ [ g_{\bar{\sigma}}^{(0)}]^{-1}( \omega )
[ g_{\sigma}^{(0)}]^{-1} ( \omega )
- Y^2 \left[
1 + \frac{ \gamma_\sigma ( \omega )}{ g_{\sigma}^{(0)} ( \omega ) }
+  \frac{ \gamma_{ \bar{\sigma}} ( \omega )}{ g_{\bar{\sigma}}^{(0)}
( \omega ) }  \right] + Y^4 \gamma_\sigma ( \omega ) \gamma_{\bar{\sigma}}
( \omega ) \right\}
\:\:\:\: .
\label{c7}
\eneq
\noindent
 Eq.(\ref{c7}) has been used in section 4 to compute the Josephson
currents near the FFP. If $1 < g < 9 /4$, for  $\gamma =( 2 k + 1 ) \pi / 3$,
${\bf O}$ is the relevant   W-instanton operator while, for $\gamma  \neq
( 2 k + 1 ) \pi / 3$,  ${\bf O}$ is the irrelevant  V-instanton operator.
In the latter situation, the $ O (Y^2)$-approximation to Eq.(\ref{c7})
allows to compute   the smoothening induced by the V-instantons
on  the sawtooth behavior of the Josephson current.
Setting, for instance,  $- \frac{\pi}{3} < \gamma < \frac{\pi}{3}$,
 $\frac{ \mu_2}{2 \pi} \sim \frac{ \mu_2^*}{2 \pi} =
- \frac{1}{\sqrt{3}}$, and  $-\frac{1}{6} < \frac{ \mu_1}{ 2 \pi}
< \frac{1}{6}$,  the states $ | \uparrow \rangle$ and $ | \downarrow
\rangle$  belong both to  sublattice A, and
${\bf O} =  - 2 Y : \cos \left[ 2 \sqrt{ \frac{2}{ 3}} \psi_2 ( 0 ) \right]:$.
In this case, Eq.(\ref{c7}) may  be approximated as

\beq
{\bf Z}_{\sigma \sigma } ( \omega ) \approx \frac{ i ( \omega - \omega_0 )
+ E_{ \bar{\sigma}} ( \vec{ \beta } ) }{ [  i ( \omega - \omega_0 )
+ E_{ \sigma} ( \vec{ \beta } ) ][  i ( \omega - \omega_0 )
+ E_{ \bar{\bar{\sigma}}} ( \vec{ \beta } ) ] - Y^2 }
\:\:\:\: .
\label{jss5}
\eneq
\noindent
The partition function is then given by

\beq
{\bf Z} [ \mu_1 , \mu_2^*   ] \approx 2  \exp \left\{ 
-  \beta \frac{g \pi v }{L}
\left[  \left( \frac{ \mu_1}{ 2 \pi} \right)^2 +
\left( \frac{ \mu_2 - \mu_2^*}{2 \pi} \right)^2 + \frac{1}{3}  -
 \sqrt{ \left( \frac{ \mu_2 -
\mu_2^*}{2 \pi} \right)^2 + \frac{y^2}{2 \pi v g} } \right] \right\}
\: ,
\label{jsc10}
\eneq
\noindent
with $y = L Y \sim L^{1 - h_S ( g )}$.
From Eq.(\ref{jsc10}), one may derive the Josephson current
distribution in the three arms  of the YJJN

\begin{eqnarray}
I_1  &=& \frac{e^* v}{L} \left\{ \frac{1}{\sqrt{2}} \left( \frac{ \mu_1}{2
\pi} \right) + \frac{1}{ \sqrt{6}}  \left( \frac{ \mu_2 -
\mu_2^*}{2 \pi} \right) \left[ 1 + \frac{1}{2} \left[
\left( \frac{ \mu_2 - \mu_2^*}{ 2 \pi }  \right)^2 + \left(
\frac{y}{ \pi v g } \right)^2 \right]^{ - \frac{1}{2} } \right]
 \right\}
\nonumber \\
I_2  &=& \frac{e^* v}{L} \left\{ -
\frac{1}{\sqrt{2}} \left( \frac{ \mu_1}{2
\pi} \right) + \frac{1}{ \sqrt{6}}
\left( \frac{ \mu_2 - \mu_2^*}{2 \pi} \right)
\left[ 1 + \frac{1}{2} \left[
\left( \frac{ \mu_2 - \mu_2^*}{ 2 \pi }  \right)^2 + \left(
\frac{y}{ \pi v g } \right)^2 \right]^{ - \frac{1}{2} } \right]
 \right\}
\nonumber \\
I_3  &=& - \frac{e^* v}{L} \sqrt{\frac{2}{3}}
\left( \frac{ \mu_2 - \mu_2^*}{2 \pi} \right)
 \left[ 1 + \frac{1}{2} \left[
\left( \frac{ \mu_2 - \mu_2^*}{ 2 \pi }  \right)^2 + \left(
\frac{y}{ \pi v g } \right)^2 \right]^{ - \frac{1}{2} } \right]
\:\:\:\: .
\label{jsc11}
\end{eqnarray}
\noindent
The formalism developed in this appendix allows also to compute
the (transverse part of the)
dynamical spin susceptibility of the emerging two-level system,
$\chi_\perp ( \Omega )$. As discussed  in section 5, the
imaginary part of $\chi_\perp ( \Omega ) / \Omega$ is the place
to look at, in order to analyze the entanglement between the
system and the environmental modes.

\begin{figure}
\includegraphics*[width=1.0\linewidth]{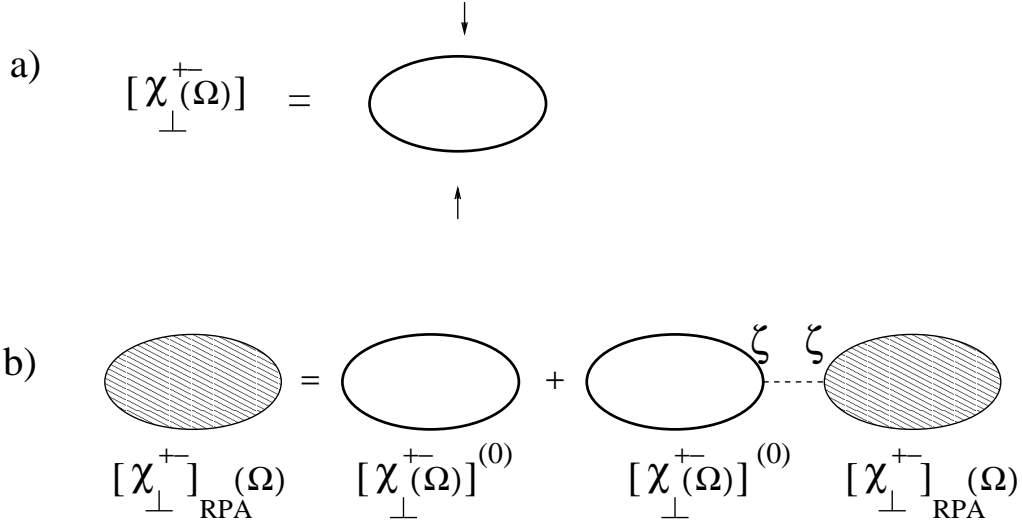}
\caption{Diagrams contributing to the  RPA computation of  the
dynamical spin susceptibility: the dashed line represents the propagation of
the W-instanton. }
\label{dyson2}
\end{figure}

The starting point to derive the $_{xx}$ and the $_{yy}$-components of
$\chi_\perp ( \Omega )$ is the computation of $\chi_\perp^{+-}
( \Omega )$ and $\chi_\perp^{-+} ( \Omega )$, that is, of
 the Fourier transforms of the imaginary time dynamical
susceptibilities $\chi_\perp^{+-} ( \tau )$, and
 $\chi_\perp^{-+} ( \tau )$, respectively given by

\beq
\chi_\perp^{+-} ( \tau ) = \langle {\bf T}_\tau [ \sigma^+ ( \tau ) \sigma^- (
0 ) ] \rangle \;\;\; , \;\;
\chi_\perp^{-+} ( \tau ) = \langle {\bf T}_\tau [ \sigma^- ( \tau ) \sigma^+ (
0 ) ] \rangle \;\;\;\: .
\label{d1}
\eneq
\noindent
The approximate computation of  $\chi_\perp^{+-} ( \Omega )$ is
graphically shown in Fig.\ref{dyson2}{\bf a)}.
To lowest order in $Y$,  $\chi_\perp^{+-} ( \Omega )$ is computed as
a loop defined by the $| \uparrow \rangle$-state propagating forward in
(imaginary) time, and by  the $| \downarrow \rangle $-state propagating
backward, while  $\chi_\perp^{-+} ( \Omega )$ is computed in the same
way, by just exchanging $\uparrow$ and $\downarrow$. Accordingly,
 $\chi_\perp^{+-} ( \Omega ) $ and $\chi_\perp^{-+} ( \Omega )$, are given by

\beq
[ \chi_\perp^{+-} ]^{(0)}
 (  i \Omega ) \approx \int \: \frac{ d \omega}{ 2 \pi }
\: {\bf Z}_{ \uparrow \uparrow } (  \omega  ) {\bf Z}_{ \downarrow \downarrow}
( \omega + \Omega )  \; , \;
[ \chi_\perp^{-+} ]^{(0)}
 (  i \Omega ) \approx \int \: \frac{ d \omega}{ 2 \pi }
\: {\bf Z}_{ \uparrow \uparrow } (  \omega + \Omega )
{\bf Z}_{ \downarrow \downarrow} ( \omega )
\: ,
\label{d2}
\eneq
\noindent
where the functions  ${\bf Z}_{ \sigma \sigma } (  \omega ) $ have
been defined in Eq.(\ref{jss5}). As a result, one obtains

\[
[ \chi_\perp^{+-} ]^{(0)}  (  i \Omega ) =
\frac{ i \Omega [ \Delta ( \vec{\mu} ) + E_\downarrow ( \vec{\mu} )] -
Y^2}{ 4 \Delta^2 ( \vec{\mu} )}
\frac{1}{ i \Omega - 2 \Delta ( \vec{\mu})}
\]
\beq
+ \frac{ i \Omega [ \Delta ( \vec{\mu} ) + E_\uparrow ( \vec{\mu} ) ] -
Y^2}{ 4 \Delta^2 ( \vec{\mu} )}
\frac{1}{ i \Omega + 2 \Delta ( \vec{\mu})}
\:\:\:\: ,
\label{d3}
\eneq
\noindent
with $\Delta ( \vec{\mu} ) = \sqrt{ [ E_\uparrow ( \vec{\mu} ) ]^2
+ Y^2 }$. A similar formula holds for $[ \chi_\perp^{+-} ]^{(0)}
(  i \Omega )$, provided one exchanges $\uparrow$ with $\downarrow$, and
vice versa,  in Eq.(\ref{d3}). For $1 < g < 9 / 4$ and $\gamma \neq
( 2 k + 1 ) \pi / 3$, and for $g > 9 / 4$, one may safely  neglect
higher order  corrections in $Y$ to $\chi_\perp ( \Omega )$, so
that  Eq.(\ref{d3}) provides a reliable estimate
of the transverse dynamical spin susceptibility. Both the $_{xx}$
and the $_{yy}$ components of $\chi_\perp ( \Omega )$ are
obtained from Eq.(\ref{d3}), and from the analogous one for $\chi_\perp^{-+}
( \Omega )$. Their imaginary part is computed
 via the replacement $\Omega \longrightarrow - i \Omega + 0^+$: in
both cases it is equal to $\chi^{``}_\perp ( \Omega )$, given by

\beq
\frac{ \chi^{``}_\perp ( \Omega ) }{ \Omega }
  \propto \delta ( \Omega - 2 \Delta ( \vec{\mu} ) )
+ \delta ( \Omega + 2 \Delta ( \vec{\beta } ) )
\:\:\:\: .
\label{dd3a}
\eneq
\noindent
Eq.(\ref{dd3a}) is the estimate of $\chi^{``}_\perp ( \Omega ) / \Omega$ near
the SFP, quoted  in section 5.1.

For $g <1$, the instantons provide a relevant perturbation: thus,
higher-order contribution in $Y$ to Eq.(\ref{d2}) cannot be neglected.
By taking the large-$Y$ limit of  the fully dressed expression for
${\bf Z}_{\sigma \sigma} ( \omega )$
derived in Eq.(\ref{c7}),  one gets, for the imaginary part of the
transverse dynamical spin susceptibility

\beq
\frac{\chi^{``}_\perp
( \Omega )}{ \Omega } \propto [ | 2 E_\uparrow ( \vec{\mu})
 + \Omega |^{3 - \frac{16}{9} g } - | 2 E_\uparrow  (\vec{\mu}) -
 \Omega |^{3 - \frac{16}{9} g}]/\Omega
\:\:\:\: .
\label{dd4}
\eneq
\noindent
Eq.(\ref{dd4}) shows that the  largest part of the spectral weight is
now in the region around $\Omega = 0$: this signals the onset of
a fully entangled state between the two state system and the bath formed by the plasmon modes
\cite{giusonew,novais}.

For $1 < g < 9 / 4$ and $\gamma = ( 2 k + 1 ) \pi / 3$,
the behavior of the system is ruled by the IR stable
FFP. An estimate of $\chi_\perp^{+-}
( \Omega )$ may now be done, for instance, when $g = \frac{9}{4} - \epsilon$,
with $\epsilon \ll 1$: since the FFP
 is at $\zeta_* \sim \epsilon^\frac{1}{2}$, using the effective
Hamiltonian in Eq.(\ref{cici2}), one may resort
to the RPA computation of the dynamical spin susceptibility, graphically
drawn in Fig.\ref{dyson2}{\bf b)}, to get

\[
[ \chi_\perp^{+-} ]_{\rm RPA}  ( \Omega ) \approx \frac{1}{ \Omega -
\Delta_* ( \vec{\mu} ) - \zeta^2
\Gamma [ - 1 - \frac{8}{9}  \epsilon ]
( - \Omega )^{ 1 + \frac{8}{9}  \epsilon } }
\]
\beq
 + \frac{1}{ \Omega + \Delta_*  ( \vec{\beta })  -
\zeta^2 \Gamma [ - 1 - \frac{8}{9}  \epsilon ]
( - \Omega )^{ 1 + \frac{8}{9}  \epsilon } }
\:\:\:\: ,
\label{ii1}
\eneq
\noindent
with $\Delta_* ( \vec{\mu} ) = \sqrt{ [ E_\uparrow ( \vec{ \beta }) ]^2
+(  \zeta_* / L)^2}$.
Computing   $\chi^{``} ( \Omega ) / \Omega$ From Eq.(\ref{ii1}), one
sees that, on one hand,
the energies of the two-level quantum system are renormalized
by $\zeta$ to $\pm \Delta_*  ( \vec{\mu}  )$,
on the other hand, that the two peaks at the renormalized energies
now display a  finite width $\propto \frac{ \pi v}{L}
( \zeta_* )^{ 1 + \frac{8}{9} g}$, which is the result quoted in section 5.2.

\begin{thebibliography}{99}



\bibitem{aoc} C. Chamon, M. Oshikawa and I. Affleck, Phys. Rev.
Lett. {\bf 91}, (2003), 206403; M. Oshikawa, C. Chamon and I.
Affleck, Journal of Statistical Mechanics JSTAT/2006/P02008;
Chang-Yu Hou and Claudio Chamon, Phys. Rev. {\bf B 77}, (2008), 155422.


\bibitem{Demler} A. Tokuno, M. Oshikawa and E. Demler, Phys. Rev. Lett. 100,
(2008), 140402.

\bibitem{doucosen} K. Kazymyrenko and B. Dou\c{c}ot, Phys. Rev.
{\bf B 71}, (2005), 075110; K. Kazymyrenko, S. Dusuel,
and B. Dou\c{c}ot, Phys. Rev.  {\bf B 72}, (2005), 235114;
S. Lal, S. Rao and D. Sen, Phys. Rev. {\bf B 66} (2002), 165327;
S. Das, S. Rao and D. Sen, Phys. Rev. {\bf B 70} (2004), 085318;
Phys. Rev. {\bf B 74} (2006), 045322;
V. R. Chandra, S. Rao, and D. Sen,  Phys. Rev.  {\bf B 75},
(2007), 045435; S. Das and S. Rao, arXiv preprint/0807.0804. 


\bibitem{guowhite} H. Guo and S. R. White,
Phys. Rev.  {\bf B 74}, (2006), 060401(R).

\bibitem{reyes} S. A. Reyes and A. M. Tsvelik, Phys. Rev.
{\bf 95}, (2005), 186404.

\bibitem{tscho} A. M. Tsvelik and P. B. Wiegmann, Adv. Phys. {\bf 32},
(1983), 453; P. Schlottmann, Phys. Rep. {\bf 181}, (1989), 1;
P. Nozieres and A. Blandin, J. Phys. (France), {\bf 41},
(1980), 193; I. Affleck and A. W. Ludwig, Phys. Rev. {\bf B 48},
(1993), 7297.


\bibitem{ager} S. Eggert and I. Affleck, Phys. Rev. {\bf B 46},
(1992), 10866.


\bibitem{fencha} P. Fendley. A. W. W. Ludwig, and H. Saleur,
 Phys. Rev. Lett. {\bf 74},  (1995), 3005; A. M. Chang, Rev. Mod. Phys.
{\bf 75}, (2003), 1449.


\bibitem{gogolin} See, for instance,
A. O. Gogolin, A. A. Nersesyan, and A. M. Tsvelik,
{\it Bosonization and Strongly Correlated Systems}, Cambridge University
Press. (2004).


\bibitem{kanefish}  C.L. Kane and M. P. Fisher, Phys. Rev. Lett. {\bf
68}, (1992), 1220; Phys. Rev. {\bf B46}, (1992), 15233.


\bibitem{giuso1} D. Giuliano and P. Sodano, Nucl. Phys. {\bf B 711},
(2005), 480.

\bibitem{giuso2} D. Giuliano and P. Sodano, Nucl. Phys. {\bf B 770},
(2007), 332.


\bibitem{glark}  L. I. Glazman and A. I. Larkin,  Phys. Rev. Lett. 79,
3736-3739 (1997).


\bibitem{glhek} F. W. J. Hekking and L. I. Glazman, Phys. Rev. {\bf B 55},
(1997), 6551.

\bibitem{kaneyi} H.Yi and C.L.Kane, Phys.Rev.{\bf B 57},R5579-R5582(1998).

\bibitem{saleurb} I. Affleck, M. Oshikawa and H.
Saleur, Nucl. Phys. {\bf B594}, (2001), 535.


\bibitem{shon0} Y. Makhlin, G. Sh\"on, and A. Shnirman, Rev. Mod.
Phys. {\bf 73}, (2001), 357.


\bibitem{swolff} J. R. Schrieffer and P. A. Wolff, Phys. Rev. {\bf 149},
                 (1966), 491.

\bibitem{luttinger} J. M. Luttinger, J. Math. Phys. {\bf 4}, (1963) 1154;
S. Tomonaga, Prog. Theor. Phys. {\bf 5}, (1950) 544.



\bibitem{pginsp} See, for example,
                P. Ginsparg, ``Applied Conformal Field Theory'', in {\it
                 Field, Strings and Critical Phenomena}, Les Houches, Section
                XLIX, (1988), Edited by E. Br\'ezin and P. Zinn-Justin.

\bibitem{giusonew} D. Giuliano and P. Sodano, New Journal of Physics {\bf 10}, 
                  (2008) 093023.



\bibitem{novais} E. Novais, A. H. Castro Neto, L. Borda, I. Affleck, and
G. Zarand, Phys. Rev. {\bf B 72},  (2005), 014417.



\bibitem{granata} C. Granata, A. Vettoliere and M. Russo, Appl. Phys.
Lett. {\bf 88}, (2006) 212506.

\bibitem{haviland} D. B. Haviland, K. Andersson, and P. Agren, J. Low
Temp. Phys.  {\bf 124},(2001)  291.

\bibitem{blatter} M. V. Feigel'man, L. B. Ioffe, V. B. Geshkenbern, P. Dayal,
and G. Blatter,  Phys. Rev. Lett {\bf 92}, 098301 (2004);  Phys.
Rev. {\bf B 70},  224524 (2004).


\bibitem{fubini} S. Fubini and G. Veneziano, Nuovo Cimento {\bf A67},
                 29 (1970); Ann. Phys. {\bf 63}, 12 (1970).


\end{thebibliography}
\end{document}